\documentclass[12pt, reqno]{amsart}

\usepackage[a4paper, margin=1.in, foot=.3in]{geometry}

\usepackage{color}

\definecolor{darkblue}{rgb}{0,0,.8}
\definecolor{darkred}{rgb}{.8,0,0}
\definecolor{darkgreen}{rgb}{0,0.8,0}

\usepackage{hyperref}
\hypersetup{
    pdfstartview=FitH,
    colorlinks,
    citecolor=darkgreen,
    linkcolor=darkred,
    urlcolor=darkblue
}

\usepackage[noBBpl,slantedGreek]{mathpazo}

\allowdisplaybreaks
\numberwithin{equation}{section}

\makeatletter
\def\subsection{\@startsection{subsection}{2}%
  \z@{.5\linespacing\@plus.7\linespacing}{-.5em}%
  {\normalfont\bfseries\mathversion{bold}}}
\makeatother

\def \qbinom #1#2{\begin{bmatrix} #1 \\ #2 \end{bmatrix}}

\def \bbC {\mathbb C}
\def \bbN {\mathbb N}
\def \bbZ {\mathbb Z}
\def \calA {\mathcal A}
\def \calB {\mathcal B}
\def \calC {\mathcal C}
\def \calD {\mathcal D}
\def \calK {\mathcal K}

\def \calR {\mathcal R}
\def \gothb {\mathfrak b}
\def \gothg {\mathfrak g}
\def \gothh {\mathfrak h}

\def \gothsl {\mathfrak{sl}}
\def \Osc {\mathrm{Osc}}
\def \rme {\mathrm e}

\def \ad {\mathrm{ad}}

\def \End {\mathrm{End}}
\def \id {\mathrm{id}}

\begin{document}

\title[Exercises with the universal $R$-matrix]{Exercises with the universal $\mathbold{R}$-matrix}

\author[H. Boos]{Herman Boos}
\address{Fachbereich C -- Physik, Bergische Universit\"at Wuppertal, 42097 Wuppertal, Germany}
\email{boos@physik.uni-wuppertal.de}

\author[F. G\"ohmann]{Frank G\"ohmann}
\address{Fachbereich C -- Physik, Bergische Universit\"at Wuppertal, 42097 Wuppertal, Germany}
\email{goehmann@physik.uni-wuppertal.de}

\author[A. Kl\"umper]{Andreas Kl\"umper}
\address{Fachbereich C -- Physik, Bergische Universit\"at Wuppertal, 42097 Wuppertal, Germany}
\email{kluemper@uni-wuppertal.de}

\author[Kh. S. Nirov]{\vskip .2em Khazret S. Nirov}
\address{Institute for Nuclear Research of the Russian Academy of Sciences, 60th October Ave 7a,
117312 Moscow, Russia}
\curraddr{Fachbereich C -- Physik, Bergische Universit\"at Wuppertal, 42097 Wuppertal, Germany}
\email{knirov@physik.uni-wuppertal.de}

\author[A. V. Razumov]{Alexander V. Razumov}
\address{Institute for High Energy Physics, 142281 Protvino, Moscow region, Russia}
\email{Alexander.Razumov@ihep.ru}

\begin{abstract}
Using the formula for the universal $R$-matrix proposed by Khoroshkin and Tolstoy, we give a
detailed derivation of $L$-operators for the quantum groups associated with the generalized Cartan
matrices $A_1^{(1)}$ and $A_2^{(1)}$.
\end{abstract}


\maketitle

\tableofcontents

\section{Introduction}

The famous statistical ice model was solved by Lieb \cite{Lie67a, Lie67b} based on a modification
of the Bethe Ansatz \cite{Bet31}. It is a special case of the more general six-vertex model which
is accessible to the Bethe Ansatz as well \cite{Sut67}. In distinction, its further generalization
to the eight-vertex model needs a different treatment proposed by Baxter \cite{Bax71, Bax72}. It is
based on the concept of the so-called $Q$-operator which, together with the transfer-matrix of the
model, satisfies a functional equation which serves as a substitute of the Bethe equations. The
$Q$-operator method has many applications in the theory of quantum integrable systems. We would
like to mention here its important role in the recent investigation of the correlation functions of
the XXZ-spin chain \cite{BooJimMiwSmiTak07, BooJimMiwSmiTak09, JimMiwSmi09, BooJimMiwSmi09}. It is
worth noting that, in general, for a given integrable system one has a set of transfer matrices and
$Q$-operators which satisfy a whole system of functional relations.

The modern approach to quantum integrable systems is based on the notion of a quantum group
introduced independently by Drinfeld \cite{Dri85, Dri87} and Jimbo \cite{Jim85, Jim86b}. In such an
approach the main object is the universal $R$-matrix which is an element in the tensor product of
two copies of the underlying quantum group $\calA$. Following Bazhanov, Lukyanov and Zamolodchikov
\cite{BazLukZam96, BazLukZam97, BazLukZam99}, one can obtain the $Q$-operators from the universal
$R$-matrix. The first step is to define the so-called $L$-operators. To this end, one realizes one
of the factors of the tensor product $\calA \otimes \calA$\footnote{To be specific, the universal
$R$-matrix in the case under consideration is an element of $\calB_+ \otimes \calB_- \subset \calA
\otimes \calA$, where $\calB_+$ and $\calB_-$ are two dual Borel subalgebras of $\calA$, while the
whole quantum group $\calA$ can be realized by means of the so-called quantum double construction
\cite{Dri87}, see also \cite{Bur90}.} in a representation space of a tensor product of
$q$-oscillator algebras, and for the remaining factor uses an appropriate finite- or
infinite-dimensional representation of $\calA$. It is common to call the representation space of
the tensor product of $q$-oscillator algebras the auxiliary space, and the remaining representation
space the quantum space. {}From the $L$-operators one constructs monodromy-type operators and takes
the trace over the $q$-oscillator factor.

For the six-vertex model and the related XXZ-spin chain the underlying quantum group is
$U_\hbar(\gothg(A_1^{(1)}))$.\footnote{We denote by $\gothg(A)$ the Kac--Moody Lie algebra defined
by the generalized Cartan matrix $A$.} This paper is devoted to the construction of $L$-operators
for quantum integrable systems which are related to the quantum group $U_\hbar(\gothg(A_2^{(1)}))$
with a finite-dimensional quantum space.\footnote{Some quantum integrable systems related to the
quantum groups $U_\hbar(\gothg(A_1^{(1)}))$ and $U_\hbar(\gothg(A_2^{(1)}))$ with
infinite-dimensional quantum spaces were studied in the papers \cite{BazLukZam96, BazLukZam97,
BazLukZam99, BazHibKho02}.}  We consider this as the first step to a generalization of the results
on correlation functions obtained in the papers \cite{BooJimMiwSmiTak07, BooJimMiwSmiTak09,
JimMiwSmi09, BooJimMiwSmi09} to the case of quantum chain related to the Lie algebra $\gothsl(3,
\bbC) \simeq \gothg(A_2)$ in the same sense as the XXZ-spin chain is related to the Lie algebra
$\gothsl(2, \bbC) \simeq \gothg(A_1)$. We start our construction directly with the universal
$R$-matrix.

An explicit formula for the universal $R$-matrix for the quantum groups $U_\hbar(\gothg(A_n))$ was
obtained in \cite{Ros89}, and for the case of the quantum groups associated with the
finite-dimensional simple Lie algebras in the \cite{KirRes90, LevSoi90}. The case of quantum groups
associated with finite-dimensional Lie superalgebras was considered in \cite{KhoTol91}. A useful
formula for the universal $R$-matrix for the quantum groups associated with the untwisted affine
Lie algebras was given first by Khoroshkin and Tolstoy \cite{TolKho92, KhoTol92}. Then this formula
was also obtained with the help of the quantum Weyl group for the case of the quantum group
$U_\hbar(\gothg(A_1^{(1)}))$ in \cite{LevSoiStu93} and for the case of the quantum groups
associated with untwisted affine Lie algebras in \cite{Dam98}. In the present paper we follow the
approach by Khoroshkin and Tolstoy \cite{TolKho92, KhoTol92}.

We start our work with the quantum group $U_\hbar(\gothg(A_1^{(1)}))$. Although the $L$-operators
for this case are well known, we believe that the analysis of this simple case provides the
necessary experience to attack the more intricate case of the quantum group
$U_\hbar(\gothg(A_{2}^{(1)}))$. Deriving $L$-operators from the universal $R$-matrix has the
advantage that one obtains them with the proper normalization implied by the functional relations
in their universal form arising when they are derived from the universal $R$-matrix as well, see,
for example, \cite{BazTsu08}. Moreover, the freedom in the construction of $L$-operators becomes
more transparent if one starts with the universal $R$-matrix. First of all, one should choose which
factor in the tensor product $\calA \otimes \calA$ will be realized with the help of $q$-oscillator
algebras. This gives two types of $L$-operators which we call $L$-operators of type $\hat L$ and
$\check L$. Furthermore, there is a freedom related to the automorphism group of the corresponding
Dynkin diagram. For the case of the affine Lie algebra $\gothg(A_1^{(1)})$ this is the symmetric
group $\mathrm{S}_2$. It has two elements, and we have two nonequivalent $L$-operators of each
type. We give explicit expressions for all of them.

In the case of the quantum group $U_\hbar(\gothg(A_2^{(1)}))$ there is an additional freedom. It
originates from the fact that realizing the generators of the quantum group by $q$-oscillators
there are two nonequivalent ways to satisfy the Serre relations. We are going to investigate a
possible manifestation of this freedom on the level of functional relations. The automorphism group
of the Dynkin diagram of the affine Lie algebra $\gothg(A_2^{(1)})$ is the dihedral group
$\mathrm{D}_3$ which coincides with the symmetric group $\mathrm{S}_3$. The order of these groups
is $6$. We do not give expressions for all possible $L$-operators restricting ourselves to two
examples related to different ways to satisfy the Serre relations for each type of $L$-operators.
We hope that the explanations given in the text are enough to obtain the remaining $L$-operators
related to the automorphism group of the Dynkin diagram.

To make the presentation more self-contained we supplemented the paper by two appendices containing
the necessary definitions of Kac--Moody algebras of finite and affine type, quantum groups and the
Khoroshkin--Tolstoy construction of the universal $R$-matrix. Further information can be found, for
example, in the books \cite{Kac90, Car05, ChaPre95, EtiFreKir98}.

We denote by $\bbZ$ the set of integers, by $\bbZ_+$ the set of positive integers, by $\bbZ_-$ the
set of negative integers, and by $\bbN$ the set of non-negative integers. We use one and the same
notation for an endomorphism of a vector space and for its matrix with respect to a fixed basis.
Hence, it is natural to denote the Kronecker product of matrices by the symbol of tensor product.
Depending on the context, the symbol $1$ means the unit of an algebra or the unit matrix.

\section{\texorpdfstring{Generalized Cartan matrix $A_1^{(1)}$}{Generalized Cartan matrix A1(1)}}

\subsection{\texorpdfstring{Universal $R$-matrix}{Universal R-matrix}}

In the case of the quantum group $U_\hbar(\gothg'(A_1^{(1)}))$, the Khoroshkin--Tolstoy
construction of the universal $R$-matrix, described in Appendix \ref{ktc}, looks as follows.

We have two simple positive roots $\alpha_0$ and $\alpha_1$. The generalized Cartan matrix
$A_1^{(1)}$ has the form
\[
A_1^{(1)} = \left( \begin{array}{rr}
2 & -2 \\
-2 & 2
\end{array} \right),
\]
therefore, $(\alpha_i, \alpha_i) = 2$ and $(\alpha_i, \alpha_j) = -2$ for $i \ne j$, see Appendix
\ref{a:A.3}. Denote $\delta = \alpha_0 + \alpha_1$ and $\alpha = \alpha_1$. It is easy to verify
that
\[
(\delta, \delta) = 0, \qquad (\delta, \alpha) = (\alpha, \delta) = 0, \qquad (\alpha, \alpha) = 2.
\]
The system of positive roots of $\gothg(A_1^{(1)})$ is
\[
\Delta_+(A_1^{(1)}) = \{ \alpha + m \delta \mid m \in \bbN \} \cup \{ m \delta \mid m \in \bbZ_+ \}
\cup \{ (\delta - \alpha) + m \delta \mid m \in \bbN \},
\]
see Appendix \ref{a:A.4}. We define the root vectors, corresponding to the roots of
$\gothg(A_1^{(1)})$, in terms of the root vectors $e_\alpha$ and $e_{\delta - \alpha}$,
corresponding to the simple positive roots, and the root vectors $f_\alpha$ and $f_{\delta -
\alpha}$, corresponding to the simple negative roots. First, using the relations
(\ref{e3})--(\ref{e6}), we define the root vectors $e_{\alpha + m \delta}$, $e_{(\delta - \alpha) +
m \delta}$ and $e'_{m \delta}$, and then, using (\ref{e7}), we introduce the root vectors $e_{m
\delta}$.\footnote{For simplicity, we denote the root vectors $e'_{m \delta, \, \alpha}$ and $e_{m
\delta, \, \alpha}$  by $e'_{m \delta}$ and $e_{m \delta}$. A similar convention is used for the
root vectors $f'_{m \delta, \, \alpha}$ and $f_{m \delta, \, \alpha}$.} Finally, using (\ref{f1})
and (\ref{f2}), we define the root vectors $f'_{m \delta}$, $f_{m \delta}$, $f_{\alpha + m \delta}$
and $f_{(\delta - \alpha) + m \delta}$.

We fix the following {\em normal order\/} of $\Delta_+(A_1^{(1)})$:
\[
\alpha, \, \alpha + \delta, \, \ldots, \, \alpha + m \delta, \, \ldots, \, \delta, \, 2 \delta, \,
\ldots, \, k \delta, \, \ldots, \, \ldots, \, (\delta - \alpha) + \ell \delta, \, \ldots, \,
(\delta - \alpha) + \delta, \, \delta - \alpha.
\]
This order evidently satisfies condition (\ref{noc}).

The expression for the universal $R$-matrix, obtained by Khoroshkin and Tolstoy, has the form
\[
\calR = \calR_{\prec \delta} \, \calR_{\sim \delta} \, \calR_{\succ \delta} \, \calK,
\]
see Appendix \ref{ktc}. The first factor is the product over $m \in \bbN$ of the $q$-exponentials
$\calR_{m, \, \alpha}$, defined by the relation (\ref{rgm}), in the order coinciding with the
chosen normal order of the roots $\alpha + m \delta$. One can verify that
\[
[e_{\alpha + m \delta}, \, f_{\alpha + m \delta}] = \frac{q^{h_{\alpha + m \delta}} - q^{-h_{\alpha
+ m \delta}}}{q - q^{-1}}.
\]
Therefore, in the case under consideration the $q$-exponentials $\calR_{m, \, \alpha}$ have the
form
\begin{equation}
\calR_{\alpha, \, m} = \exp_{q^{-2}} \left( (q - q^{-1}) \, e_{\alpha + m
\delta} \otimes f_{\alpha + m \delta} \right). \label{rma}
\end{equation}
The matrices $u_m$, whose matrix elements enter the expression (\ref{rpd}) for $\calR_{\sim
\delta}$, are just numbers, and we have
\begin{equation}
\calR_{\sim \delta} = \exp \left( (q - q^{-1}) \sum_{m \in \bbZ_+} \frac{m}{[2 m]_q} e_{m \delta}
\otimes f_{m \delta} \right). \label{rpda1}
\end{equation}
The factor $\calR_{\succ \delta}$ is the product over $m \in \bbN$ of the $q$-exponentials
$\calR_{\delta - \alpha, \, m}$, defined by the relation
(\ref{rdmgm}), in the order coinciding with the chosen normal order of the roots $(\delta - \alpha)
+ m \delta$. Since
\[
[e_{(\delta - \alpha) + m \delta}, \, f_{(\delta - \alpha) + m \delta}]
= \frac{q^{h_{(\delta - \alpha) + m \delta}} - q^{-h_{(\delta - \alpha) + m \delta}}}{q - q^{-1}},
\]
the $q$-exponentials $\calR_{\delta - \alpha, \, m}$ have the form
\begin{equation}
\calR_{\delta - \alpha, \, m} = \exp_{q^{-2}} \left( (q -
q^{-1}) \, e_{(\delta - \alpha) + m \delta} \otimes f_{(\delta - \alpha) + m \delta} \right).
\label{rmdma1}
\end{equation}
Finally, for the last factor $\calK$ we have
\begin{equation}
\calK = \exp(\hbar \, h_\alpha \otimes h_\alpha / 2),
\label{k1}
\end{equation}
see equation (\ref{k}).

\subsection{\texorpdfstring{$R$-matrix. Fundamental representation}{R-matrix. Fundamental representation}}

As a warm-up exercise we will reproduce the $R$-matrix corresponding to the two-dimensional
representations of the quantum group $U_\hbar(\gothg'(A_1^{(1)}))$, determined by the first
fundamental representation of the quantum group $U_\hbar(\gothg(A_1))$. Some formulas of this
section are necessary for the construction of $L$-operators.

First, we define a homomorphism $\varepsilon$ from $U_\hbar(\gothg'(A_1^{(1)}))$ to $U_\hbar(\gothg(A_1))$ by
its action on the generators \cite{Jim86a}:\footnote{We denote the generators $h_\alpha$, $e_\alpha$ and $f_\alpha$
of the quantum group $U_\hbar(\gothg(A_1))$ by $H$, $E$ and $F$ respectively.}
\begin{align*}
\varepsilon \, (h_{\alpha_0}) = \varepsilon \, (h_{\delta - \alpha})
= - H, \qquad & \varepsilon \, (h_{\alpha_1}) = \varepsilon \, (h_\alpha) = H, \\
\varepsilon \, (e_{\alpha_0}) = \varepsilon \, (e_{\delta - \alpha})
= F, \qquad & \varepsilon \, (e_{\alpha_1}) = \varepsilon \, (e_\alpha) = E, \\
\varepsilon \, (f_{\alpha_0}) = \varepsilon \, (f_{\delta - \alpha})
= E, \qquad & \varepsilon \, (f_{\alpha_1}) = \varepsilon \, (f_\alpha) = F.
\end{align*}
It is not difficult to verify that our definition is consistent with the defining relations of the
quantum group $U_\hbar(\gothg'(A_1^{(1)}))$. Note that the Serre relations here have the form
\begin{gather*}
e_i^3 \, e_j - [3]_q \, e_i^2 \, e_j \, e_i + [3]_q \, e_i \, e_j \, e_i^2 - e_j \, e_i^3 = 0, \\
f_i^3 \, f_j - [3]_q \, f_i^2 \, f_j \, f_i + [3]_q \, f_i \, f_j \, f_i^2 - f_j \, f_i^3 = 0.
\end{gather*}

The first and the only fundamental representation $\pi^{(1)}$ of $U_\hbar(\gothg(A_1))$ is two dimensional and
coincides with the first fundamental representation of the Lie algebra $\gothg(A_1)$. Hence, we have
\[
\pi^{(1)}(H) = E_{11} - E_{22}, \qquad
\pi^{(1)}(E) = E_{12}, \qquad
\pi^{(1)}(F) = E_{21}.
\]
We define a homomorphism $\varphi$ as
\[
\varphi = \pi^{(1)} \circ \varepsilon,
\]
and then, with the help of equations (\ref{phiz1}) and (\ref{phiz2}), define the homomorphisms
$\varphi_\zeta$. It is clear that they can be explicitly defined by the equations\footnote{In
formulas related to the case of the generalized Cartan matrix $A_1^{(1)}$ we use the integers $s =
s_0 + s_1$ and $s_1$ instead of $s_0$ and $s_1$.}
\begin{align}
\varphi_\zeta(h_{\delta - \alpha}) = - E_{11} + E_{22}, \qquad &\varphi_\zeta(h_\alpha)
= E_{11} - E_{22},
\label{vha} \\
\varphi_\zeta(e_{\delta - \alpha}) = \zeta^{s - s_1} E_{21}, \qquad &\varphi_\zeta(e_\alpha)
= \zeta^{s_1} E_{12},
\label{pi1e} \\
\varphi_\zeta(f_{\delta - \alpha}) = \zeta^{- s + s_1} E_{12}, \qquad &\varphi_\zeta(f_\alpha)
= \zeta^{-s_1} E_{21}.
\label{pi1f}
\end{align}
It is worth noting that if we have an expression for $\varphi_\zeta(a)$, where $a$ is an element of
$U_\hbar(\gothg'(A_1^{(1)}))$, then to obtain the expression for $\varphi_\zeta(\omega(a))$, where
$\omega$ is the Cartan anti-involution, defined by equations (\ref{ci}), one can simply take the
transpose of $\varphi_\zeta(a)$ and change $q$ to $q^{-1}$ and $\zeta$ to $\zeta^{-1}$. We observe
that $\varphi_{\zeta_1} \otimes \varphi_{\zeta_2} (\calR)$ depends only on $\zeta_{12} =
\zeta_1/\zeta_2$, and so we define
\[
R(\zeta_{12}) = \varphi_{\zeta_1} \otimes \varphi_{\zeta_2} (\calR).
\]

At this point it is possible to obtain the expression for $\varphi_{\zeta_1} \otimes
\varphi_{\zeta_2} (\calK)$. Indeed, it follows from equation (\ref{vha}) that
\begin{multline*}
\varphi_{\zeta_1} \otimes \varphi_{\zeta_2} (h_\alpha \otimes h_\alpha)
= (E_{11} - E_{22}) \otimes (E_{11} - E_{22}) \\
= E_{11} \otimes E_{11} - E_{11} \otimes E_{22} - E_{22} \otimes E_{11} + E_{22} \otimes E_{22},
\end{multline*}
and, taking into account (\ref{k1}), we obtain
\begin{multline}
\varphi_{\zeta_1} \otimes \varphi_{\zeta_2} (\calK) = \varphi_{\zeta_1} \otimes \varphi_{\zeta_2}
(\exp(\hbar \, h_\alpha \otimes h_\alpha /2)) \\*
= q^{1/2} E_{11} \otimes E_{11} + q^{-1/2} E_{11} \otimes E_{22} + q^{-1/2} E_{22}
\otimes E_{11} + q^{1/2} E_{22} \otimes E_{22}.
\label{fka1}
\end{multline}

To find the expression for $\varphi_{\zeta_1} \otimes \varphi_{\zeta_2} (\calR_{\prec \delta})$ and
$\varphi_{\zeta_1} \otimes \varphi_{\zeta_2} (\calR_{\succ \delta})$ we need expressions for
$\varphi_\zeta(e_{\alpha + m \delta})$, $\varphi_\zeta(f_{\alpha + m \delta})$,
$\varphi_\zeta(e_{(\delta - \alpha) + m \delta})$, and $\varphi_\zeta(f_{(\delta - \alpha) + m
\delta})$. We start with equations (\ref{pi1e}). It follows from (\ref{e3}) that
\[
\varphi_\zeta(e'_\delta) = \zeta^s (E_{11} - q^{-2} E_{22}),
\]
and the recursive definitions (\ref{e4}) and (\ref{e5}) give
\begin{gather}
\varphi_\zeta(e_{\alpha + m\delta}) = (-1)^m q^{-m} \zeta^{s_1 + m s} E_{12},
\label{eapmd} \\
\varphi_\zeta(e_{(\delta - \alpha) + m\delta}) = (-1)^m q^{-m} \zeta^{(s - s_1) + m s} E_{21}.
\label{edmapmd}
\end{gather}
Now, using the relation (\ref{f2}), we obtain
\begin{gather}
\varphi_\zeta(f_{\alpha + m\delta}) = (-1)^m q^m \zeta^{- s_1 - m s} E_{21},
\label{fapmd} \\
\varphi_\zeta(f_{(\delta - \alpha) + m\delta}) = (-1)^m q^m \zeta^{- (s - s_1) - m s} E_{12}.
\label{fdmapmd}
\end{gather}
It follows from equations (\ref{eapmd}) and (\ref{fapmd}) that
\[
\varphi_{\zeta_1} \otimes \varphi_{\zeta_2} (e_{\alpha + m \delta} \otimes f_{\alpha + m \delta})
= \zeta^{s_1 + m s}_{12} E_{12} \otimes E_{21}.
\]
Taking into account the identities
\begin{equation}
(E_{12})^k = 0, \qquad (E_{21})^k = 0
\label{eek}
\end{equation}
valid for any integer $k > 1$, we find
\[
\varphi_{\zeta_1} \otimes \varphi_{\zeta_2} \left( \calR_{\alpha, \, m}
\right) = 1 + (q - q^{-1}) \zeta^{s_1 + m s}_{12} E_{12} \otimes E_{21},
\]
where $\calR_{m, \, \alpha}$ is given by (\ref{rma}). Using again (\ref{eek}), we obtain
\begin{multline}
\varphi_{\zeta_1} \otimes \varphi_{\zeta_2} ( \calR_{\prec \delta})
= 1 + (q - q^{-1}) \sum_{m \in \bbN} \zeta^{s_1 + m s}_{12} E_{12} \otimes E_{21} \\*
= 1 + (q - q^{-1}) \frac{\zeta^{s_1}_{12}}{1 - \zeta^s_{12}} \, E_{12} \otimes E_{21}.
\label{frlda1}
\end{multline}
In a similar way, starting from (\ref{edmapmd}) and (\ref{fdmapmd}) and taking into account
(\ref{rmdma1}), we derive the equation
\begin{equation}
\varphi_{\zeta_1} \otimes \varphi_{\zeta_2} ( \calR_{\succ \delta})
= 1 + (q - q^{-1}) \frac{\zeta^{s - s_1}_{12}}{1 - \zeta^s_{12}} \, E_{21} \otimes E_{12}.
\label{frgda1}
\end{equation}

We will find now the expression for $\varphi_{\zeta_1} \otimes \varphi_{\zeta_2} ( \calR_{\sim \delta})$.
It follows from (\ref{e6}) and (\ref{eapmd}) that
\[
\varphi_\zeta(e'_{m \delta}) = (-1)^{m - 1} q^{-m + 1} \zeta^{m s} (E_{11} - q^{-2} E_{22}).
\]
Hence, we have
\[
\varphi_\zeta(e'_\delta(x)) = \frac{q \, \zeta^s}{q x + \zeta^s} (E_{11} - q^{-2} E_{22}),
\]
where $e'_\delta(x) = \sum_{m \in \bbZ_+} e'_{m \delta} x^{-m}$. Simple calculations give
\[
\varphi_\zeta(\log(1 + (q - q^{-1}) e'_\delta(x)))
= \log \frac{1 + q \, \zeta^s x^{-1}}{1 + q^{-1} \zeta^s x^{-1}} \, E_{11} + \log \frac{1 + q^{-3}
\zeta^s x^{-1}}{1 + q^{-1} \zeta^s x^{-1}} \, E_{22}.
\]
In our case equation (\ref{e7}) has the form
\[
(q - q^{-1}) e_\delta(x) = \log(1 + (q - q^{-1}) e'_\delta(x)),
\]
where $e_\delta(x) = \sum_{m \in \bbZ_+} e_{m \delta} x^{-m}$, and we obtain
\begin{equation}
\varphi_\zeta(e_{m\delta}) = (-1)^{m-1} \frac{[m]_q}{m} \, \zeta^{m s} (E_{11} - q^{-2m} E_{22}).
\label{emd}
\end{equation}
Having in mind that $f_{m \delta} = \omega(e_{m \delta})$, we determine that
\begin{equation}
\varphi_\zeta(f_{m\delta}) = (-1)^{m-1} \frac{[m]_q}{m} \, \zeta^{- m s} (E_{11} - q^{2m} E_{22}).
\label{fmd}
\end{equation}
Equations (\ref{emd}) and (\ref{fmd}) give
\begin{multline*}
\varphi_{\zeta_1} \otimes \varphi_{\zeta_2} \Bigl( (q - q^{-1}) \sum_{m \in \bbZ_+}
\frac{m}{[2 m]_q} e_{m \delta} \otimes f_{m \delta} \Bigr) \\*
= \sum_{m \in \bbZ_+} \, \frac{q^m - q^{-m}}{q^m + q^{-m}} \, \frac{\zeta^{m s}_{12}}{m} (E_{11}
\otimes E_{11} - q^{2m} E_{11} \otimes E_{22} - q^{-2m} E_{22} \otimes E_{11} + E_{22} \otimes E_{22}).
\end{multline*}
Introduce the function
\begin{equation}
\lambda_2(\zeta) = \sum_{m \in \bbZ_+} \, \frac{1}{q^m + q^{-m}} \, \frac{\zeta^m}{m}
= \sum_{m \in \bbZ_+} \, \frac{1}{[2]_{q^m}} \, \frac{\zeta^m}{m},
\label{fa1}
\end{equation}
then, we can write
\begin{multline*}
\varphi_{\zeta_1} \otimes \varphi_{\zeta_2} \Bigl( (q - q^{-1})
\sum_{m \in \bbZ_+} \frac{m}{[2 m]_q} e_{m \delta} \otimes f_{m \delta} \Bigr) \\*
= [\lambda_2(q \zeta^s_{12}) - \lambda_2(q^{-1} \zeta^s_{12})]E_{11} \otimes E_{11}
+ [\lambda_2(q \zeta^s_{12}) - \lambda_2(q^3 \zeta^s_{12})] E_{11} \otimes E_{22} \\*[.5em]
+ [\lambda_2(q^{-3} \zeta^s_{12}) - \lambda_2(q^{-1} \zeta^s_{12})] E_{22} \otimes E_{11}
+ [\lambda_2(q \zeta^s_{12}) - \lambda_2(q^{-1} \zeta^s_{12})] E_{22} \otimes E_{22}.
\end{multline*}
Taking into account (\ref{rpda1}), we obtain
\begin{multline*}
\varphi_{\zeta_1} \otimes \varphi_{\zeta_2} (\calR_{\sim \delta}) =
\rme^{\lambda_2(q \zeta^s_{12}) - \lambda_2(q^{-1} \zeta^s_{12})} \Bigl[ E_{11} \otimes E_{11} \\*
+ \rme^{\lambda_2(q^{-1} \zeta^s_{12}) - \lambda_2(q^3 \zeta^s_{12})} E_{11} \otimes E_{22}
+ \rme^{\lambda_2(q^{-3} \zeta^s_{12}) - \lambda_2(q \zeta^s_{12})} E_{22}
\otimes E_{11} + E_{22} \otimes E_{22} \Bigr].
\end{multline*}
It is easy to determine that
\[
\lambda_2(q \zeta) + \lambda_2(q^{-1} \zeta) = - \log (1 - \zeta).
\]
Therefore, we have
\begin{multline}
\varphi_{\zeta_1} \otimes \varphi_{\zeta_2} (\calR_{\sim \delta}) =
\rme^{\lambda_2(q \zeta^s_{12}) - \lambda_2(q^{-1} \zeta^s_{12})} \biggl[ E_{11} \otimes E_{11}
\\+ \frac{1 - q^2 \zeta^s_{12}}{1 - \zeta^s_{12}} \, E_{11} \otimes E_{22}
+ \frac{1 - \zeta^s_{12}}{1 - q^{-2} \zeta^s_{12}} \, E_{22} \otimes E_{11} + E_{22} \otimes E_{22} \biggr].
\label{frpda1}
\end{multline}

Now we have the expressions (\ref{frlda1}), (\ref{frgda1}), (\ref{frpda1}) and (\ref{fka1}) for all
factors necessary to obtain the expression for $R(\zeta)$. After simple calculations we determine
that
\begin{multline}
R(\zeta) = q^{1/2} \rme^{\lambda_2(q \zeta^s) - \lambda_2(q^{-1} \zeta^s)} \biggl[ E_{11} \otimes E_{11}
+ E_{22} \otimes E_{22} + \frac{q^{-1} (1 - \zeta^s)}{1 - q^{-2} \zeta^s} (E_{11} \otimes E_{22} \\*
+ E_{22} \otimes E_{11}) + \frac{1 - q^{-2}}{1 - q^{-2} \zeta^s} (\zeta^{s_1} E_{12}
\otimes E_{21} + \zeta^{s - s_1} E_{21} \otimes E_{12}) \biggr].
\label{rm2}
\end{multline}
We come to the most frequently used symmetric $R$-matrix putting $s = -2$ and $s_1 = -1$.

As was noted in the paper \cite{BraGouZhaDel94}, $R$-matrices corresponding to different values of
$s$ and $s_1$ are connected by a change of the spectral parameters and a gauge transformation.
Actually, one can convince oneself that
\begin{equation}
R^{(s, s_1)}(\zeta_{12}) = [G(\zeta_1) \otimes G(\zeta_2)]R^{(1, 0)}(\zeta^s_{12})[G(\zeta_1)
\otimes G(\zeta_2)]^{-1}, \label{gt2}
\end{equation}
where
\begin{equation}
G(\zeta) = \left( \begin{array}{cc}
1 & 0 \\
0 & \zeta^{-s_1}
\end{array} \right)
\label{gz}
\end{equation}
and the dependence on $s$ and $s_1$ is shown explicitly.\footnote{It is implied here that $R(\zeta)
= R^{(s, s_1)}(\zeta)$, $R^{(1,0)}(\zeta^s) = R^{(s,0)}(\zeta)$, and a similar convention is also
assumed for the corresponding $L$-operators.} The relation (\ref{gt2}) is more general than the
relation given in the paper \cite{BraGouZhaDel94}. It can be used to show that transfer-matrices of
{\em inhomogeneous\/} vertex models corresponding to $R$-matrices with different choice of $s$ and
$s_1$ are connected by a similarity transformation and a simple change of the spectral parameters.
The corresponding partition functions in the case of toroidal boundary conditions are connected by
a change of the spectral parameters.

\subsection{\texorpdfstring{$L$-operators. Oscillator algebra}{L-operator. Oscillator algebra}}

In this section, to construct $L$-operators we will use the $q$-oscillator algebra $\Osc_\hbar$
defined as an associative algebra with generators $a$, $a^\dagger$, $D$ and relations\footnote{As
usual, we consider $\hbar$ as an indeterminate, assume that $q = \exp \hbar$, and treat the
elements of $\Osc_\hbar$ as formal power series in $\hbar$.}
\begin{align*}
[D, a] = -a, \qquad & [D, a^\dagger] = a^\dagger, \\* a \, a^\dagger = 1 - q^2 q^{2D},
\qquad & a^\dagger a = 1 - q^{2D}.
\end{align*}
We should define homomorphisms $\varphi_\zeta$ and $\chi_\zeta$ or $\psi_\zeta$, see Appendix
\ref{a:B.6}. In this section we always define homomorphisms $\varphi_\zeta$  by equations
(\ref{vha})--(\ref{pi1f}).

The transformations
\begin{equation}
a \to \kappa \, a \, q^{\xi D}, \qquad a^\dagger \to \frac{1}{\kappa} \, q^{- \xi D} a^\dagger, \qquad D \to D
\label{oa}
\end{equation}
form a two-parameter group of automorphisms of the algebra $\Osc_\hbar$. One can apply a transformation of
this group to an $L$-operator and obtain another $L$-operator. The trace used to define $Q$-operators in the
case under consideration is invariant with respect to the transformations~(\ref{oa}), see, for example,
\cite{BazLukZam97, BooJimMiwSmiTak07}. Therefore, the $Q$-opera\-tors obtained from $L$-operators connected
by the transformations (\ref{oa}) coincide. Thus, we can call $L$-operators connected by the transformations
(\ref{oa}) equivalent.

To construct $L$-operators of type $\hat L$ one can use homomorphisms $\chi$ defined by the relations
\begin{gather}
\chi(h_{\delta - \alpha}) = - 2 D, \qquad \chi(h_\alpha) = 2 D, \label{chih} \\
\chi(e_{\delta - \alpha}) = \rho \, \mu \, a \, q^{\nu D}, \qquad \chi(e_{\alpha})
= \frac{1}{\mu} \, q^{- \nu D} a^\dagger, \label{chie}
\end{gather}
where $\rho$, $\mu$ and $\nu$ are free parameters. One can verify that the above definition is consistent with the
defining relations of the algebra $U_\hbar(\gothb'_+(A_1^{(1)}))$.
The parameters $\mu$ and $\nu$ in the final expressions can be freely changed by the transformations (\ref{oa}).
Hence, the $L$-operators corresponding to different values of $\mu$ and $\nu$ are equivalent. Changing the parameter
$\rho$ we change the coefficient at $\zeta^s$ in the final expression.

To construct $L$-operators of type $\check L$ one can use homomorphisms $\psi$ defined by the relations
\begin{gather}
\psi(h_{\delta - \alpha}) = - 2 D, \qquad \psi(h_\alpha) = 2 D, \label{psih} \\
\psi(e_{\delta - \alpha}) = \rho \, \frac{1}{\mu} \, q^{- \nu D} a^\dagger, \qquad \psi(e_{\alpha})
= \mu \, a \, q^{\nu D}, \label{psif}
\end{gather}
where again $\rho$, $\mu$ and $\nu$ are free parameters. The parameters $\mu$ and $\nu$ in the final expressions can
be freely changed by the transformations (\ref{oa}), and changing the parameter $\rho$ we change the coefficient at
$\zeta^s$ in the final expression.

There are a few methods to obtain $L$-operators which are not equivalent to those which can be obtained by using the
homomorphisms defined by the relations (\ref{chih}), (\ref{chie}) and (\ref{psih}), (\ref{psif}).

First of all, recall that there is a standard automorphism $\sigma$ of the quantum group
$U_\hbar(\gothg'(A_1^{(1)}))$, corresponding to the only automorphism of the Dynkin diagram of the
generalized Cartan matrix $A_1^{(1)}$. It is defined by the relations
\begin{gather}
\sigma(h_{\alpha_0}) = h_{\alpha_1}, \qquad \sigma(h_{\alpha_1}) = h_{\alpha_0}, \label{sig_h}\\
\sigma(e_{\alpha_0}) = e_{\alpha_1}, \qquad \sigma(e_{\alpha_1}) = e_{\alpha_0}, \qquad
\sigma(f_{\alpha_0}) = f_{\alpha_1}, \qquad \sigma(f_{\alpha_1}) = f_{\alpha_0}. \label{sig_ef}
\end{gather}
Applying first the automorphism $\sigma$, we modify the homomorphisms $\chi$ and $\psi$ defined by
the relations (\ref{chih}), (\ref{chie}) or (\ref{psih}), (\ref{psif}) and come to $L$-operators
which are not equivalent to those which are obtained by applying the unmodified
homomorphisms (\ref{chih}), (\ref{chie}) or (\ref{psih}), (\ref{psif}).

Further, it is not difficult to show that if $\hat L(\zeta)$ is an $L$-operator of type $\hat L$, then
the $L$-operator $\hat L^{-1}(\zeta^{-1})$ is of type $\check L$, and if $\check L(\zeta)$ is an $L$-operator
of type $\check L$, then $\check L^{-1}(\zeta^{-1})$ is an $L$-operator of type $\hat L$.

Finally, it is easy to determine that the mapping $\tau$, defined by the equations
\begin{equation}
\tau(a) = a^\dagger, \qquad \tau(a^\dagger) = a, \qquad \tau(D) = D,
\label{oaa}
\end{equation}
is an anti-involution of the algebra $\Osc_\hbar$. Then one can be convinced that if $\hat
L(\zeta)$ is an $L$-operator of type $\hat L$, then $\tau(\hat L(\zeta^{-1}))$ is an $L$-operator
of type $\check L$, and vice versa, if $\check L(\zeta)$ is an $L$-operator of type $\check L$,
then $\tau(\check L(\zeta^{-1}))$ is an $L$-operator of type $\hat L$.

\subsubsection{Type $\hat L$}

We define a homomorphism $\chi$ from $U_\hbar(\gothb'_+(A_1^{(1)}))$ to $\Osc_\hbar$ by the
relations (\ref{chih}), (\ref{chie}) with $\rho = 1 / (q - q^{-1})^2$, $\mu = q - q^{-1}$ and $\nu
= 0$ so that
\begin{align}
\chi(h_{\delta - \alpha}) = -2D, \qquad & \chi(h_\alpha) = 2D,
\label{ch2} \\
\chi(e_{\delta - \alpha}) = \frac{1}{q - q^{-1}} \, a, \qquad & \chi(e_\alpha) = \frac{1}{q - q^{-1}} \, a^\dagger.
\label{ce2}
\end{align}
The corresponding homomorphisms $\chi_\zeta$, $\zeta \in \bbC^\times$, are defined with the help of
the procedure described at the end of Appendix \ref{ktc}. We have explicitly
\begin{align}
\chi_\zeta(h_{\delta - \alpha}) = -2D, \qquad & \chi_\zeta(h_\alpha) = 2D,
\label{ch1} \\
\chi_\zeta(e_{\delta - \alpha}) = \frac{1}{q - q^{-1}} \, a \, \zeta^{s - s_1}, \qquad & \chi_\zeta(e_\alpha)
= \frac{1}{q - q^{-1}} \, a^\dagger \, \zeta^{s_1}.
\label{ce1}
\end{align}
Here $\chi_{\zeta_1} \otimes \varphi_{\zeta_2}(\calR)$ depends on $\zeta_{12}$ only, and we define
$\hat L(\zeta)$ by equation (\ref{hlz}).

Having in mind (\ref{k1}), (\ref{vha}) and (\ref{ch1}), we observe that
\begin{equation}
\chi_{\zeta_1} \otimes \varphi_{\zeta_2}(\calK) = q^D E_{11} + q^{-D} E_{22}.
\label{ok1}
\end{equation}
Further, one can easily determine that the definition (\ref{e3}) together with (\ref{ce1}) gives
\begin{equation}
\chi_\zeta(e'_\delta) = \frac{1}{q - q^{-1}} \, q^{-1} \zeta^s,
\label{cepd}
\end{equation}
and, using (\ref{e4}), we immediately obtain
\begin{equation}
\chi_\zeta(e_{\alpha + m \delta}) = 0, \qquad \chi_\zeta(e_{(\delta - \alpha) + m \delta}) = 0, \qquad m \ge 1.
\label{ceamd1}
\end{equation}
Taking into account (\ref{fapmd}), (\ref{fdmapmd}) and (\ref{eek}), we come to
\begin{equation}
\chi_{\zeta_1} \otimes \varphi_{\zeta_2}(\calR_{\prec \delta}) = 1 + a^\dagger \zeta^{s_1}_{12} E_{21},
\qquad \chi_{\zeta_1} \otimes \varphi_{\zeta_2}(\calR_{\succ \delta}) = 1 + a \, \zeta^{s - s_1}_{12} E_{12}.
\label{orld1}
\end{equation}
The definition (\ref{e6}) and equations (\ref{ceamd1}) give
\[
\chi_\zeta(e'_{m \delta}) = 0, \qquad m \ge 1,
\]
and one easily finds that
\[
\chi_\zeta(e_{m \delta}) = \frac{1}{q - q^{-1}} \, (-1)^{m-1} q^{-m} \, \frac{\zeta^{ms}}{m}.
\]
Now, using the relations (\ref{rpda1}) and (\ref{fmd}), we obtain
\begin{equation}
\chi_{\zeta_1} \otimes \varphi_{\zeta_2}(\calR_{\sim \delta})
= \rme^{\lambda_2(q^{-1} \zeta^s_{12})} [E_{11} + (1 - \zeta^s_{12}) E_{22}],
\label{orpd1}
\end{equation}
where the function $\lambda_2(\zeta)$ is defined by (\ref{fa1}).

Multiplying the expressions (\ref{ok1}), (\ref{orld1}) and (\ref{orpd1}) in the corresponding
order, we come to the following $L$-operator:
\[
\hat L(\zeta) = \rme^{\lambda_2(q^{-1} \zeta^s)} [q^D E_{11} + a \, q^{-D} \zeta^{s - s_1} E_{12}
+ a^\dagger q^D \zeta^{s_1} E_{21} + (q^{-D} - q^D \zeta^s) E_{22}].
\]
In the matrix form it looks as
\[
\hat L(\zeta) = \rme^{\lambda_2(q^{-1} \zeta^s)} \left( \begin{array}{cc}
q^D & a \, q^{-D} \zeta^{s - s_1} \\[.5em]
a^\dagger q^D \zeta^{s_1} & q^{-D} - q^D \zeta^s
\end{array} \right).
\]

It is evident that $L$-operators corresponding to different values of $s$ and $s_1$ are connected
by a change of the spectral parameter and a gauge transformation via the equation
\[
\hat L^{(s, s_1)}(\zeta_{12}) = \gamma_{\zeta_1}(G(\zeta_2) \hat
L^{(1,0)}(\zeta^s_{12})G^{-1}(\zeta_2)).
\]
Here the matrix $G(\zeta)$ is given by equation (\ref{gz}) and the mapping $\gamma_\zeta:
\Osc_\hbar \to \Osc_\hbar$ is defined by the relations
\begin{equation}
\gamma_\zeta(a) = a \, \zeta^{- s_1}, \qquad \gamma_\zeta(a^\dagger) = a^\dagger \zeta^{s_1},
\qquad \gamma_\zeta(D) = D. \label{sgz}
\end{equation}
For any $\zeta \in \bbC^\times$ the mapping $\gamma_\zeta$ is an automorphism of the algebra
$\Osc_\hbar$ of the type defined by the relations (\ref{oa}). As we noted above the trace used to
define $Q$-operators in the case under consideration is invariant with respect to the
transformations~(\ref{oa}). Therefore, the $Q$-operators obtained from $L$-operators corresponding
to different values of $s$ and $s_1$ are connected by a change of the spectral parameters and a
similarity transformation.

Applying the automorphism $\sigma$ defined by equations\footnote{Actually we use the restriction of
the automorphism $\sigma$ to $U_\hbar(\gothb'_+(A_1^{(1)}))$.} (\ref{sig_h}), (\ref{sig_ef}) and
using the homomorphism $\chi$ defined by equations (\ref{ch2}) and (\ref{ce2}), according to the
procedure described at the end of Appendix \ref{ktc} we obtain
\begin{align*}
\chi_\zeta(h_{\delta - \alpha}) = 2D, \qquad & \chi_\zeta(h_\alpha) = -2D, \\* \chi_\zeta(e_{\delta
- \alpha}) = \frac{1}{q - q^{-1}} \, a^\dagger \, \zeta^{s - s_1}, \qquad & \chi_\zeta(e_\alpha) =
\frac{1}{q - q^{-1}} \, a \, \zeta^{s_1}.
\end{align*}
This leads to the following $L$-operator
\[
\hat L(\zeta) = \rme^{\lambda_2(q^{-1} \zeta^s)} \left( \begin{array}{cc}
q^{-D} - q^D \zeta^s & a^\dagger q^D \zeta^{s - s_1} \\[.5em]
a \, q^{-D} \zeta^{s_1} & q^D
\end{array} \right).
\]

\subsubsection{Type $\check L$} The calculations needed to obtain $L$-operators of type
$\check L$ start with construction of a homomorphism $\psi$ from $U_\hbar(\gothb'_+(A_1^{(1)}))$ to
$\Osc_\hbar$. We define it using the relations (\ref{psih}), (\ref{psif}) with $\rho = 1 / (q -
q^{-1})^2$, $\mu = 1/ (q - q^{-1})$ and $\nu = 0$ so that
\begin{align}
\psi(h_{\delta - \alpha}) = -2D, \qquad & \psi(h_\alpha) = 2D,
\label{ph} \\
\psi(f_{\delta - \alpha}) = \frac{1}{q - q^{-1}} \, a^\dagger, \qquad & \psi(f_\alpha) = \frac{1}{q - q^{-1}} \, a.
\label{pef}
\end{align}
Using the procedure of Appendix \ref{ktc}, we come to the $\zeta$-dependent homomorphisms
$\psi_\zeta$ for which
\begin{align*}
\psi_\zeta(h_{\delta - \alpha}) = -2D, \qquad & \psi_\zeta(h_\alpha) = 2D, \\
\psi_\zeta(f_{\delta - \alpha}) = \frac{1}{q - q^{-1}} \, a^\dagger \zeta^{-(s - s_1)},
\qquad & \psi_\zeta(f_\alpha) = \frac{1}{q - q^{-1}} \, a \, \zeta^{-s_1}.
\end{align*}
Here $\varphi_{\zeta_1} \otimes \psi_{\zeta_2}(\calR)$ depend on $\zeta_{12}$ only and we define
$\check L(\zeta)$ by equation (\ref{clz}).

As above, it is easy to show that
\begin{equation}
\varphi_{\zeta_1} \otimes \psi_{\zeta_2}(\calK) = q^D E_{11} + q^{-D} E_{22}.
\label{ok2}
\end{equation}
Using equation (\ref{f6}), we determine that
\[
\psi_\zeta(f'_\delta) = \frac{1}{q - q^{-1}} \, q^2 \, [(q + q^{-1}) q^{2D} - q^{-1}] \, \zeta^{-s}.
\]
This allows, taking into account (\ref{f4}) and (\ref{f5}), to obtain the following expressions:
\begin{gather}
\psi_\zeta(f_{\alpha + m \delta}) = \frac{1}{q - q^{-1}} \, (-1)^m q^m a \, q^{2mD} \zeta^{-s_1 -
ms},
\label{pzfamd} \\
\psi_\zeta(f_{(\delta - \alpha) + m \delta}) = \frac{1}{q - q^{-1}} \, (-1)^m q^m q^{2mD} a^\dagger
\zeta^{- (s - s_1) - ms}. \label{psfdamd}
\end{gather}
Using these expressions, we come to the equations
\begin{gather}
\varphi_{\zeta_1} \otimes \psi_{\zeta_2}(\calR_{\prec \delta})
= 1 + a \, (1 - q^{2D} \zeta^s_{12})^{-1} \zeta^{s_1}_{12} E_{12},
\label{orld2} \\*
\varphi_{\zeta_1} \otimes \psi_{\zeta_2}(\calR_{\succ \delta})
= 1 + (1 - q^{2D} \zeta^s_{12})^{-1} a^\dagger \zeta^{s - s_1}_{12} E_{21}.
\label{orgd2}
\end{gather}
With the account of (\ref{pzfamd}) and (\ref{psfdamd}), it follows from (\ref{f6}) that
\begin{multline*}
\psi_\zeta(f'_{m \delta}) = \frac{1}{(q - q^{-1})^2} \, (-1)^{m-1} q^{2m} \\*
\times [(q^{m+1} - q^{-m-1}) q^{2mD} - q^{-1}(q^m - q^{-m}) q^{2(m-1)D}] \, \zeta^{-ms}.
\end{multline*}
Using the relation (\ref{f7}), we obtain
\[
\psi_\zeta(f_{m \delta}) = \frac{1}{q - q^{-1}} \, (-1)^m q^m \, [1 - (1 + q^{2m}) q^{2mD}] \,
\frac{\zeta^{-ms}}{m}.
\]
This leads to the equation
\begin{equation}
\varphi_{\zeta_1} \otimes \psi_{\zeta_2}(\calR_{\sim \delta})
= \rme^{\lambda_2(q^{-1} \zeta^s_{12})} \, [(1 - \zeta^s_{12}) (1 - q^2 q^{2D} \zeta^s_{12})^{-1}
E_{11} + (1 - q^{2D} \zeta^s_{12}) E_{22}].
\label{orpd2}
\end{equation}

After all, using the expressions (\ref{orld2}), (\ref{orpd2}), (\ref{orgd2}) and (\ref{ok2}),
we obtain the following $L$-operator
\[
\check L(\zeta) = \rme^{\lambda_2(q^{-1} \zeta^s)} \, [q^D E_{11} + a \, q^{-D} \zeta^{s_1} E_{12}
+ a^\dagger q^D \zeta^{s - s_1} E_{21} + (q^{-D} - q^D \zeta^s)E_{22}]
\]
with the matrix form
\[
\check L(\zeta) = \rme^{\lambda_2(q^{-1} \zeta^s)} \left( \begin{array}{cc}
q^D & a \, q^{-D} \zeta^{s_1} \\[.5em]
a^\dagger q^D \zeta^{s - s_1} & q^{-D} - q^D \zeta^s
\end{array} \right).
\]

To connect $L$-operators corresponding to different values of $s$ and $s_1$ one can use the
equation
\[
\check L^{(s, s_1)}(\zeta_{12}) = G(\zeta_1) \gamma_{\zeta_2}(\check L^{(1,0)}(\zeta^s_{12}))
G^{-1}(\zeta_1),
\]
where the matrix $G(\zeta)$ is defined by equation (\ref{gz}) and the mapping $\gamma_\zeta$ by the
relations (\ref{sgz}).

Using the restriction of the automorphism $\sigma$ defined by equations (\ref{sig_h}),
(\ref{sig_ef}) to the subalgebra $U_\hbar(\gothb'_-(A_1^{(1)}))$ and then applying the homomorphism
$\psi$ defined by equations (\ref{ph}) and (\ref{pef}) we can proceed to the $\zeta$-dependent
homomorphisms $\psi_\zeta$, such that
\begin{gather*}
\psi_\zeta(h_{\delta - \alpha}) = 2D, \qquad \psi_\zeta(h_\alpha) = -2D, \\
\psi_\zeta(f_{\delta - \alpha}) = \frac{1}{q - q^{-1}} \, a \, \zeta^{-(s - s_1)}, \qquad
\psi_\zeta(f_\alpha) = \frac{1}{q - q^{-1}} \, a^\dagger \zeta^{-s_1}.
\end{gather*}
In this way we come to the $L$-operator
\[
\check L(\zeta) = \rme^{\lambda_2(q^{-1} \zeta^s)} \left( \begin{array}{cc}
q^{-D} - q^D \zeta^s & a^\dagger q^D \zeta^{s_1} \\[.5em]
a \, q^{-D} \zeta^{s - s_1} & q^D
\end{array} \right).
\]

\section{\texorpdfstring{Generalized Cartan matrix $A_2^{(1)}$}{Generalized Cartan matrix A2(1)}}

\subsection{\texorpdfstring{Universal $R$-matrix}{Universal R-matrix}}

In the case of $\gothg(A_2^{(1)})$ there are three simple positive roots $\alpha_0$, $\alpha_1$ and
$\alpha_2$. The generalized Cartan matrix $A_2^{(1)}$ has the form
\[
A_2^{(1)} = \left( \begin{array}{rrr}
2 & -1 & -1 \\
-1 & 2 & -1 \\
-1 & -1 & 2
\end{array} \right),
\]
therefore, $(\alpha_i, \alpha_i) = 2$ and $(\alpha_i, \alpha_j) = -1$ for $i \ne j$, see Appendix
\ref{a:A.3}. Denote $\delta = \alpha_0 + \alpha_1 + \alpha_2$, $\alpha = \alpha_1$, and $\beta =
\alpha_2$. It is easy to verify that
\begin{gather*}
(\delta, \delta) = 0, \qquad (\delta, \alpha) = (\alpha, \delta) = 0,
\qquad (\delta, \beta) = (\beta, \delta) = 0, \\
(\alpha, \alpha) = 2, \qquad (\alpha, \beta) = (\beta, \alpha) = -1,
\qquad (\beta, \beta) = 2.
\end{gather*}

The system of positive roots $\Delta_+(A_2)$ is formed by the roots $\alpha$, $\beta$ and $\alpha +
\beta$. The system of positive roots of $\gothg(A_2^{(1)})$ is
\begin{multline*}
\Delta_+(A_2^{(1)}) = \{ \gamma + m \delta \mid \gamma \in \Delta_+(A_2), \, m \in \bbN \} \\
\cup \{ m \delta \mid m \in \bbZ_+ \} \cup \{ (\delta - \gamma) + m \delta \mid \gamma \in
\Delta_+(A_2), \, m \in \bbN \},
\end{multline*}
see Appendix \ref{a:A.4}. We define the root vectors, corresponding to the roots of
$\gothg(A_2^{(1)})$ in terms of the root vectors $e_\alpha$, $e_\beta$ and $e_{\delta - \alpha -
\beta}$, corresponding to the simple positive roots, and the root vectors $f_\alpha$, $f_\beta$ and
$f_{\delta - \alpha - \beta}$, corresponding to the simple negative roots.

First, using the definitions (\ref{e1}) and (\ref{e2}), we construct the root vector corresponding
to the root $\alpha + \beta$ as
\begin{equation}
e_{\alpha + \beta} = e_{\alpha} \, e_{\beta} - q^{-1} e_\beta \, e_\alpha,
\label{eab}
\end{equation}
and the root vectors corresponding to the roots $\delta - \alpha$ and $\delta - \beta$ as
\begin{equation}
e_{\delta - \alpha} = e_\beta \, e_{\delta - \alpha - \beta} - q^{-1} e_{\delta - \alpha - \beta}
\, e_\beta, \qquad e_{\delta - \beta} = e_\alpha \, e_{\delta - \alpha - \beta} - q^{-1} e_{\delta
- \alpha - \beta} \, e_\alpha. \label{eda2}
\end{equation}
Then, for any $\gamma \in \Delta_+(A_2)$, using the relations (\ref{e2})--(\ref{e6}), we define the
root vectors $e_{\gamma + m \delta}$, $e_{(\delta - \gamma) + m \delta}$ and $e'_{m \delta, \,
\gamma}$, and then, using (\ref{e7}), introduce the root vectors $e_{m \delta, \, \gamma}$.
Finally, using (\ref{f1}) and (\ref{f2}), we define the root vectors $f'_{m \delta, \, \gamma}$,
$f_{m \delta, \, \gamma}$, $f_{\gamma + m \delta}$ and $f_{(\delta - \gamma) + m \delta}$.

We fix the following normal order of $\Delta_+(A_2^{(1)})$ \cite{BraGouZha95}:
\begin{gather*}
\alpha, \, \alpha + \beta, \, \alpha + \delta, \, \alpha + \beta + \delta, \, \ldots, \, \alpha +
m_1 \delta, \, \alpha + \beta + m_1 \delta, \, \ldots, \\* \beta, \, \beta + \delta, \, \ldots, \,
\beta + m_2 \delta, \, \ldots, \delta, \, 2 \delta, \, \ldots, \, k \delta, \, \ldots, \ldots, \,
(\delta - \beta) + \ell_1 \delta, \, \ldots, \, \delta - \beta, \\* \ldots, (\delta - \alpha) +
\ell_2 \delta, \, (\delta - \alpha - \beta) + \ell_2 \delta, \, \ldots, \, \delta - \alpha, \,
\delta - \alpha - \beta.
\end{gather*}

The universal $R$-matrix $\calR$ has again the form
\[
\calR = \calR_{\prec \delta} \, \calR_{\sim \delta} \, \calR_{\succ \delta} \, \calK.
\]
The factor $\calR_{\prec \delta}$ is the product over $\gamma \in \Delta_+(A_2)$ and $m \in \bbN$
of the $q$-exponentials
\begin{equation}
\calR_{\gamma, \, m} = \exp_{q^{-2}}[(q - q^{-1}) e_{\gamma + m \delta} \otimes f_{\gamma + m \delta}],
\label{rgm2}
\end{equation}
see the relation (\ref{rgm}), in the order coinciding with the chosen normal order of the roots
$\gamma + m \delta$. The matrices $u_m$ entering the expression (\ref{rpd}) for the factor
$\calR_{\sim \delta}$ have in the case under consideration the form
\begin{multline}
u_m = \frac{m}{[m]_q} \frac{1}{[3]_{q^m}}
\left( \begin{array}{cc}
[2]_{q^m} & (-1)^m \\[.5em]
(-1)^m & [2]_{q^m}
\end{array} \right) \\*
= \frac{m}{[m]_q} \frac{1}{q^{2m} + 1 + q^{-2m}}
\left( \begin{array}{cc}
q^m + q^{-m} & (-1)^m \\[.5em]
(-1)^m & q^m + q^{-m}
\end{array} \right).
\label{um}
\end{multline}
The factor $\calR_{\succ \delta}$ is the product over $\gamma \in \Delta_+(A_2)$ and $m \in \bbN$
of the $q$-exponentials
\begin{equation}
\calR_{\delta - \gamma, \, m} = \exp_{q^{-2}}[(q - q^{-1}) e_{(\delta - \gamma) + m \delta} \otimes
f_{(\delta - \gamma) + m \delta}], \label{rdgm2}
\end{equation}
see the relation (\ref{rdmgm}), in the order coinciding with the chosen normal order of the roots
$(\delta - \gamma) + m \delta$. Finally, since
\[
A_2^{-1} = \frac{1}{3} \left(
\begin{array}{cc}
2 & 1 \\
1 & 2
\end{array}
\right),
\]
we have for the last factor $\calK$ the expression
\begin{equation}
\calK = \exp \left[ \hbar \, (2 h_\alpha \otimes h_\alpha + h_\alpha \otimes h_\beta + h_\beta
\otimes h_\alpha + 2 h_\beta \otimes h_\beta)/3 \right], \label{k2}
\end{equation}
see equation (\ref{k}).

\subsection{\texorpdfstring{$R$-matrix. First fundamental representation}{R-matrix. First fundamental representation}}

Now we use the standard homomorphism from $U_\hbar(\gothg'(A_2^{(1)}))$ to $U_\hbar(\gothg(A_2))$
which is defined by its action on the generators as follows \cite{Jim86a}:\footnote{We denote the
generators $h_{\alpha_i}$, $e_{\alpha_i}$ and $f_{\alpha_i}$ of the quantum group
$U_\hbar(\gothg(A_2))$ by $H_{\alpha_i}$, $E_{\alpha_i}$ and $F_{\alpha_i}$.}
\begin{gather*}
\varepsilon(h_{\alpha_0}) = \varepsilon(h_{\delta - \alpha - \beta}) = {} - H_\alpha - H_\beta, \\[.5em]
\varepsilon(h_{\alpha_1}) = \varepsilon(h_\alpha) = H_\alpha, \qquad \varepsilon(h_{\alpha_2})
= \varepsilon(h_\beta) = H_\beta, \\[.5em]
\varepsilon(e_{\alpha_0}) = \varepsilon(e_{\delta - \alpha - \beta})
= (F_\beta F_\alpha - q F_\alpha F_\beta) \, q^{-(H_\alpha - H_\beta)/3}, \\[.5em]
\varepsilon(e_{\alpha_1}) = \varepsilon(e_\alpha)
= E_\alpha, \qquad \varepsilon(e_{\alpha_2}) = \varepsilon(e_\beta) = E_\beta, \\[.5em]
\varepsilon(f_{\alpha_0}) = \varepsilon(f_{\delta - \alpha - \beta})
= (E_\alpha E_\beta - q^{-1} E_\beta E_\alpha) \, q^{(H_\alpha - H_\beta)/3}, \\[.5em]
\varepsilon(f_{\alpha_1}) = \varepsilon(f_\alpha)
= F_\alpha, \qquad \varepsilon(f_{\alpha_2}) = \varepsilon(f_\beta) = F_\beta.
\end{gather*}
It is possible to get convinced that this definition is consistent with the defining relations of
the quantum group $U_\hbar(\gothg'(A_2^{(1)}))$. Here the Serre relations have the form
\begin{equation}
e_i^2 \, e_j - [2]_q \, e_i \, e_j \, e_i + e_j \, e_i^2 = 0, \qquad f_i^2 \, f_j - [2]_q \, f_i \,
f_j \, f_i + f_j \, f_i^2 = 0. \label{sa2}
\end{equation}
The first fundamental representation $\pi^{(1,0)}$ of the quantum group $U_\hbar(\gothg(A_2))$ is
three dimensional and coincides with the first fundamental representation of the Lie algebra
$\gothg(A_2)$. Therefore, we have
\begin{gather*}
\pi^{(1,0)}(H_\alpha) = E_{11} - E_{22}, \qquad \pi^{(1,0)}(H_\beta) = E_{22} - E_{33}, \\[.5em]
\pi^{(1,0)}(E_\alpha) = E_{12}, \qquad \pi^{(1,0)}(E_\beta) = E_{23}, \\[.5em]
\pi^{(1,0)}(F_\alpha) = E_{21}, \qquad \pi^{(1,0)}(F_\beta) = E_{32}.
\end{gather*}
We define a homomorphism $\varphi$ as
\[
\varphi = \pi^{(1,0)} \circ \varepsilon,
\]
and then, using equations (\ref{phiz1}) and (\ref{phiz2}), define the homomorphisms
$\varphi_\zeta$. These homomorphisms can be defined directly by the equations\footnote{In formulas
related to the case of the generalized Cartan matrix $A_2^{(1)}$ we use instead of the integers
$s_0$, $s_1$ and $s_2$  the integers $s = s_0 + s_1 + s_2$, $s_1$ and $s_2$.}
\begin{gather}
\varphi_\zeta(h_{\delta - \alpha - \beta}) = {} - E_{11} - E_{33}, \qquad
\varphi_\zeta(h_\alpha) = E_{11} - E_{22}, \qquad \varphi_\zeta(h_\beta) = E_{22} - E_{33},
\label{phih2} \\[.5em]
\varphi_\zeta(e_{\delta - \alpha - \beta}) = \zeta^{s - s_1 - s_2} E_{31}, \qquad
\varphi_\zeta(e_\alpha) = \zeta^{s_1} E_{12}, \qquad \varphi_\zeta(e_\beta) = \zeta^{s_2} E_{23},
\label{phie2} \\[.5em]
\varphi_\zeta(f_{\delta - \alpha - \beta}) = \zeta^{-(s - s_1 - s_2)} E_{13}, \qquad
\varphi_\zeta(f_\alpha) = \zeta^{-s_1} E_{21}, \qquad \varphi_\zeta(f_\beta) = \zeta^{-s_2} E_{32}.
\label{phif2}
\end{gather}
We removed the factors $q^{-1/3}$ and $q^{1/3}$ in the definition of $\varphi_\zeta(e_{\delta -
\alpha - \beta})$ and $\varphi_\zeta(f_{\delta - \alpha - \beta})$, respectively. This can be done
using a simple automorphism of $U_\hbar(\gothg'(A_2^{(1)}))$. As in the previous case, if we have
an expression for $\varphi_\zeta(a)$, where $a$ is an element of $U_\hbar(\gothg'(A_2^{(1)}))$, and
then, to obtain the expression for $\varphi_\zeta(\omega(a))$, one can simply take the transpose of
$\varphi_\zeta(a)$ and change $q$ to $q^{-1}$ and $\zeta$ to $\zeta^{-1}$.

We start the calculation of the factors needed to construct $\varphi_{\zeta_1} \otimes
\varphi_{\zeta_2}(\calR)$ with the factor $\varphi_{\zeta_1} \otimes \varphi_{\zeta_2}(\calK)$.
Using equations (\ref{phih2}), we obtain
\begin{multline*}
\varphi_{\zeta_1} \otimes \varphi_{\zeta_2}(2 h_\alpha \otimes h_\alpha + h_\alpha \otimes h_\beta
+ h_\beta \otimes h_\alpha + 2 h_\beta \otimes h_\beta)
\\= 2 (E_{11} \otimes E_{11} + E_{22} \otimes E_{22}+ E_{33} \otimes E_{33}) \\
- (E_{11} \otimes E_{22} + E_{11} \otimes E_{33} + E_{22} \otimes E_{11} + E_{22} \otimes E_{33} +
E_{33} \otimes E_{11} + E_{33} \otimes E_{22}),
\end{multline*}
and the relation (\ref{k2}) gives
\begin{multline*}
\varphi_{\zeta_1} \otimes \varphi_{\zeta_2}(\calK) = q^{2/3} (E_{11} \otimes E_{11} + E_{22}
\otimes E_{22}+ E_{33} \otimes E_{33})
\\+ q^{-1/3} (E_{11} \otimes E_{22} + E_{11} \otimes E_{33} + E_{22} \otimes E_{11} + E_{22}
\otimes E_{33} + E_{33} \otimes E_{11} + E_{33} \otimes E_{22}).
\end{multline*}

It follows from (\ref{eab}) and (\ref{phie2}) that
\[
\varphi_\zeta(e_{\alpha + \beta}) = \zeta^{s_1 + s_2} E_{13}
\]
and, using (\ref{eda2}), we obtain
\begin{equation}
\varphi_\zeta(e_{\delta - \alpha}) = \zeta^{s - s_1} E_{21}, \qquad \varphi_\zeta(e_{\delta -
\beta}) = \zeta^{s - s_2} E_{32}. \label{phizda2}
\end{equation}
Taking into account (\ref{f2}), we come to the expressions
\begin{gather*}
\varphi_\zeta(f_{\alpha + \beta}) = \zeta^{- (s_1 + s_2)} E_{31}, \\[.5em]
\varphi_\zeta(f_{\delta - \alpha}) = \zeta^{- (s - s_1)} E_{12}, \qquad \varphi_\zeta(f_{\delta - \beta})
= \zeta^{- (s - s_2)} E_{23}.
\end{gather*}
Now, using the definition (\ref{e3}), we obtain
\begin{gather*}
\varphi_\zeta(e'_{\delta, \, \alpha}) = \zeta^s (E_{11} - q^{-2} E_{22}),
\qquad \varphi_\zeta(e'_{\delta, \, \beta})
= -q^{-1} \zeta^s (E_{22} - q^{-2} E_{33}), \\
\varphi_\zeta(e'_{\delta, \, \alpha + \beta}) = \zeta^s (E_{11} - q^{-2} E_{33}).
\end{gather*}
Using these relations and recalling the definitions (\ref{e4}) and (\ref{e5}), we determine that
\begin{gather}
\varphi_\zeta (e_{\alpha + m \delta}) = (-1)^m q^{-m} \zeta^{s_1 + ms} E_{12}, \qquad
\varphi_\zeta (e_{\beta + m \delta}) = q^{-2 m} \zeta^{s_2 + ms} E_{23},
\label{phizeamd2} \\[.5em]
\varphi_\zeta (e_{\alpha + \beta + m \delta}) = (-1)^m q^{-m} \zeta^{s_1 + s_2 + ms} E_{13}, \\[.5em]
\varphi_\zeta (e_{(\delta - \alpha) + m \delta}) = (-1)^m q^{-m} \zeta^{(s - s_1) + ms} E_{21}, \\[.5em]
\varphi_\zeta (e_{(\delta - \beta) + m \delta}) = - q^{- 2m - 1} \zeta^{(s - s_2) + s_1 + ms} E_{32}, \\[.5em]
\varphi_\zeta (e_{(\delta - \alpha - \beta) + m \delta}) = (-1)^m q^{-m} \zeta^{(s - s_1 - s_2) + ms} E_{31}.
\label{12}
\end{gather}
Taking into account (\ref{f2}), we obtain
\begin{gather}
\varphi_\zeta (f_{\alpha + m \delta}) = (-1)^m q^m \zeta^{- s_1 - ms} E_{21}, \qquad
\varphi_\zeta (f_{\beta + m \delta}) = q^{2 m} \zeta^{- s_2 - ms} E_{32},
\label{phizf21} \\[.5em]
\varphi_\zeta (f_{\alpha + \beta + m \delta}) = (-1)^m q^m \zeta^{- (s_1 + s_2) - ms} E_{31},
\\[.5em]
\varphi_\zeta (f_{(\delta - \alpha) + m \delta}) = (-1)^m q^m \zeta^{- (s - s_1) - ms} E_{12}, \\[.5em]
\varphi_\zeta (f_{(\delta - \beta) + m \delta}) = - q^{2m + 1} \zeta^{- (s - s_2) - ms} E_{23}, \\[.5em]
\varphi_\zeta (f_{(\delta - \alpha - \beta) + m \delta}) = (-1)^m q^m \zeta^{- (s - s_1 - s_2) - ms} E_{13}.
\label{phizf22}
\end{gather}
The above relations allow us to find the expressions for $\varphi_{\zeta_1} \otimes
\varphi_{\zeta_2}(\calR_{\prec \delta})$ and $\varphi_{\zeta_1} \otimes
\varphi_{\zeta_2}(\calR_{\succ \delta})$. First, using equation (\ref{rgm2}) and the fact that
\[
(E_{ab})^k = 0
\]
for all $a \ne b$ and any integer $k > 1$, we come to the expressions
\begin{gather*}
\varphi_{\zeta_1} \otimes \varphi_{\zeta_2}(\calR_{\alpha, \, m})
= 1 + (q - q^{-1}) \, \zeta^{s_1 + ms}_{12} \, E_{12} \otimes E_{21}, \\[.5em]
\varphi_{\zeta_1} \otimes \varphi_{\zeta_2}(\calR_{\alpha + \beta, \, m})
= 1 + (q - q^{-1}) \, \zeta^{s_1 + s_2 + ms}_{12} \, E_{13} \otimes E_{31}, \\[.5em]
\varphi_{\zeta_1} \otimes \varphi_{\zeta_2}(\calR_{\beta, \, m})
= 1 + (q - q^{-1}) \, \zeta^{s_2 + ms}_{12} \, E_{23} \otimes E_{32}.
\end{gather*}
It is easy to understand that $\varphi_{\zeta_1} \otimes \varphi_{\zeta_2}(\calR_{\alpha, \, m})$
and $\varphi_{\zeta_1} \otimes \varphi_{\zeta_2}(\calR_{\alpha + \beta, \, n})$ commute for any $m$
and $n$. Therefore, we can rearrange the factors entering $\varphi_{\zeta_1} \otimes
\varphi_{\zeta_2}(\calR_{\prec \delta})$ in such a way that the factors corresponding to the roots
$\alpha + m \delta$ come first, then the factors corresponding to the roots $\alpha + \beta + m
\delta$, and finally the factors corresponding to the roots $\beta + m \delta$. After that,
performing multiplication and summing up the arising geometrical series, we come to the expression
\begin{multline*}
\varphi_{\zeta_1} \otimes \varphi_{\zeta_2}(\calR_{\prec \delta}) \\* = 1 + (q - q^{-1}) \,
\frac{1}{1 - \zeta^s_{12}} \left[ \zeta^{s_1}_{12} \, E_{12} \otimes E_{21} + \zeta^{s_1 +
s_2}_{12} \, E_{13} \otimes E_{31} + \zeta^{s_2}_{12} \, E_{23} \otimes E_{32} \right].
\end{multline*}
In a similar way we obtain
\begin{multline*}
\varphi_{\zeta_1} \otimes \varphi_{\zeta_2}(\calR_{\succ \delta}) \\* = 1 + (q - q^{-1}) \,
\frac{1}{1 - \zeta^s_{12}} \left[ \zeta^{s - s_1}_{12} \, E_{21} \otimes E_{12} + \zeta^{s - s_1 -
s_2}_{12} \, E_{31} \otimes E_{13} + \zeta^{s - s_2}_{12} \, E_{32} \otimes E_{23} \right].
\end{multline*}

Now we will find the expression for $\varphi_{\zeta_1} \otimes \varphi_{\zeta_2}(\calR_{\sim \delta})$.
To this end we should find expressions for $\varphi_\zeta (e_{m \delta,
\, \alpha})$, $\varphi_\zeta (e_{m \delta, \, \beta})$, $\varphi_\zeta (f_{m \delta, \, \alpha})$
and $\varphi_\zeta (f_{m \delta, \, \beta})$. First, using (\ref{e6}), (\ref{phizda2}) and
(\ref{phizeamd2}), we obtain
\begin{gather*}
\varphi_\zeta (e'_{m \delta, \, \alpha})
= (-1)^{m - 1} q^{- m + 1} \zeta^{ms} (E_{11} - q^{-2} E_{22}), \\[.5em]
\varphi_\zeta (e'_{m \delta, \, \beta})
= - q^{- 2m + 1} \zeta^{ms} (E_{22} - q^{-2} E_{33}).
\end{gather*}
This gives
\begin{gather*}
\varphi_\zeta(\log(1 + (q - q^{-1}) e'_{\delta, \, \alpha}(x)))
= \log \frac{1 + q \, \zeta^s x^{-1}}{1 + q^{-1} \zeta^s x^{-1}} \, E_{11}
+ \log \frac{1 + q^{-3} \zeta^s x^{-1}}{1 + q^{-1} \zeta^s x^{-1}} \, E_{22}, \\
\varphi_\zeta(\log(1 + (q - q^{-1}) e'_{\delta, \, \beta}(x))) = \log \frac{1 - \zeta^s x^{-1}}{1 -
q^{-2} \zeta^s x^{-1}} \, E_{22} + \log \frac{1 - q^{-4} \zeta^s x^{-1}}{1 - q^{-2} \zeta^s x^{-1}}
\, E_{33},
\end{gather*}
and using equation (\ref{e7}) we come to the expressions
\begin{gather*}
\varphi_\zeta (e_{m \delta, \, \alpha})
= (-1)^{m - 1} \frac{[m]_q}{m} \zeta^{ms} (E_{11} - q^{-2m} E_{22}), \\[.5em]
\varphi_\zeta (e_{m \delta, \, \beta})
= - \frac{[m]_q}{m} q^{-m} \zeta^{ms} (E_{22} - q^{-2m} E_{33}).
\end{gather*}
Then, taking into account equations (\ref{f1}), we determine that
\begin{gather*}
\varphi_\zeta (f_{m \delta, \, \alpha})
= (-1)^{m - 1} \frac{[m]_q}{m} \zeta^{-ms} (E_{11} - q^{2m} E_{22}), \\[.5em]
\varphi_\zeta (f_{m \delta, \, \beta})
= - \frac{[m]_q}{m} q^m \zeta^{-ms} (E_{22} - q^{2m} E_{33}).
\end{gather*}
Recalling that the matrices $u_m$ in the case under consideration are given by (\ref{um}), we obtain
\begin{multline*}
\varphi_{\zeta_1} \otimes \varphi_{\zeta_2} \Bigl( (q - q^{-1}) \sum_{m \in \bbZ_+} \sum_{i, j =
1}^2 u_{m, \, ij} \, e_{m \delta, \alpha_i}
\otimes f_{m \delta, \, \alpha_j} \Bigr) \\
= [\lambda_3(q^2 \zeta^s_{12}) - \lambda_3(q^{-2} \zeta^s_{12})](E_{11}
\otimes E_{11} + E_{22} \otimes E_{22} + E_{33} \otimes E_{33}) \\[.5em]
+ [\lambda_3(q^2 \zeta^s_{12}) - \lambda_3(q^4 \zeta^s_{12})](E_{11}
\otimes E_{22} + E_{11} \otimes E_{33} + E_{22} \otimes E_{33}) \\[.5em]
+ [\lambda_3(q^{-4} \zeta^s_{12}) - \lambda_3(q^{-2} \zeta^s_{12})](E_{22}
\otimes E_{11} + E_{33} \otimes E_{11} + E_{33} \otimes E_{22}),
\end{multline*}
where the function $\lambda_3(\zeta)$ is defined as
\[
\lambda_3(\zeta) = \sum_{m \in \bbZ_+} \frac{1}{q^{2m} + 1 + q^{-2m}} \frac{\zeta^{m}}{m} = \sum_{m
\in \bbZ_+} \frac{1}{[3]_{q^m}} \frac{\zeta^{m}}{m}.
\]
It is easy to determine that
\[
\lambda_3(q^2 \zeta) + \lambda_3(\zeta) + \lambda_3(q^{-2} \zeta) = - \log (1 - \zeta).
\]
Using this identity and the definition (\ref{rpd}), we come to the expression
\begin{multline*}
\varphi_{\zeta_1} \otimes \varphi_{\zeta_2} (\calR_{\sim \delta}) = \rme^{\lambda_3(q^2
\zeta^s_{12}) - \lambda_3(q^{-2} \zeta^s_{12})}\biggl[ E_{11}
\otimes E_{11} + E_{22} \otimes E_{22} + E_{33} \otimes E_{33} \\
+ \frac{1 - q^2 \zeta^s_{12}}{1 - \zeta^s_{12}}(E_{11} \otimes E_{22} + E_{11}
\otimes E_{33} + E_{22} \otimes E_{33}) \\
+ \frac{1 - \zeta^s_{12}}{1 - q^{-2} \zeta^s_{12}}(E_{22} \otimes E_{11} + E_{33}
\otimes E_{11} + E_{33} \otimes E_{22}) \biggr].
\end{multline*}

Collecting all necessary factors, we obtain
\begin{multline*}
R(\zeta) = q^{2/3} \rme^{\lambda_3(q^2 \zeta^s)
- \lambda_3(q^{-2} \zeta^s)} \biggl[ E_{11} \otimes E_{11}
+ E_{22} \otimes E_{22} + E_{33} \otimes E_{33} \\*
+ \frac{q^{-1} (1 - \zeta^s)}{1 - q^{-2} \zeta^s}(E_{11} \otimes E_{22}
+ E_{11} \otimes E_{33} + E_{22} \otimes E_{11} + E_{22} \otimes E_{33} + E_{33} \otimes E_{11} \\*
+ E_{33} \otimes E_{22}) + \frac{1 - q^{-2}}{1 - q^{-2} \zeta^s}(\zeta^{s_1} E_{12} \otimes E_{21}
+ \zeta^{s_1 + s_2} E_{13} \otimes E_{31} + \zeta^{s_2} E_{23} \otimes E_{32}
\\*
+ \zeta^{s - s_1} E_{21} \otimes E_{12} + \zeta^{s - s_1 - s_2} E_{31} \otimes E_{13} +
\zeta^{s - s_2} E_{32} \otimes E_{23}) \biggr].
\end{multline*}
Note that, up to a scalar factor, the obtained $R$-matrix takes the well-known simple form when one
chooses $s = -2$ and $s_1 = s_2 = 0$:
\begin{multline*}
R(\zeta) = q^{2/3} \rme^{\lambda_3(q^2 \zeta^{-2}) - \lambda_3(q^{-2} \zeta^{-2})}
\biggl[ \sum_a E_{aa} \otimes E_{aa} + \frac{\zeta - \zeta^{-1}}{q \zeta - q^{-1}\zeta^{-1}}
\sum_{a \ne b} E_{aa} \otimes E_{bb}
\\+  \frac{(q - q^{-1})\zeta}{q \zeta - q^{-1}\zeta^{-1}} \sum_{a < b} E_{ab}
\otimes E_{ba} + \frac{(q - q^{-1})\zeta^{-1}}{q \zeta - q^{-1}\zeta^{-1}}
\sum_{a > b} E_{ab} \otimes E_{ba} \biggl].
\end{multline*}

One can verify that\footnote{It is implied here that $R(\zeta) = R^{(s,s_1,s_2)}(\zeta)$,
$R^{(1,0,0)}(\zeta^s) = R^{(s,0,0)}(\zeta)$, and a similar convention is to be used for the
corresponding $L$-operators.}
\[
R^{(s, s_1, s_2)}(\zeta_{12}) = [G(\zeta_1) \otimes G(\zeta_2)]R^{(1, 0,
0)}(\zeta^s_{12})[G(\zeta_1) \otimes G(\zeta_2)]^{-1},
\]
where
\begin{equation}
G(\zeta) = \left( \begin{array}{ccc}
1 & 0 & 0\\
0 & \zeta^{-s_1} & 0 \\
0 & 0 & \zeta^{- s_1 - s_2}
\end{array} \right).
\label{gz2}
\end{equation}
Therefore, as in the previous case, the transfer-matrices of inhomogeneous vertex models
corresponding to $R$-matrices with different choice of $s$, $s_1$ and $s_2$ are connected by a
similarity transformation and a change of the spectral parameters. The corresponding partition
functions in the case of the toroidal boundary conditions are connected by a change of the spectral
parameters.

\subsection{\texorpdfstring{$L$-operators. Oscillator algebra}{L-operator. Oscillator algebra}}

Here to construct $L$-operators we will use homomorphisms to the tensor product of two copies of
the $q$-oscillator algebra $\Osc_\hbar \otimes \Osc_\hbar$. As is usual, we define
\begin{gather*}
a_1 = a \otimes 1, \qquad a_1^\dagger = a ^\dagger \otimes 1,
\qquad a_2 = 1 \otimes a, \qquad a_2^\dagger = 1 \otimes a^\dagger, \\[.5em]
D_1 = D \otimes 1, \qquad D_2 = 1 \otimes D.
\end{gather*}
It is worth to note that the transformations
\begin{gather}
a_1 \to \kappa_1 \, a_1 \, q^{\xi_1 D_1 + \xi_2 D_2}, \qquad a_1^\dagger \to \frac{1}{\kappa_1} \,
q^{- \xi_1 D_1 - \xi_2 D_2} a_1^\dagger, \qquad D_1 \to D_1, \label{qoa1} \\* a_2 \to \kappa_2 \,
a_{2} \, q^{\xi_2 D_1 + \xi_3 D_2}, \qquad a_2^\dagger \to \frac{1}{\kappa_2} \, q^{- \xi_2 D_1 -
\xi_3 D_2} a_{2}^\dagger, \qquad D_2 \to D_2, \label{qoa2}
\end{gather}
form a five-parameter automorphism group of the algebra $\Osc_\hbar \otimes \Osc_\hbar$. The trace
used to define $Q$-operators in the case under consideration is invariant with respect to the
transformations~(\ref{qoa1}), (\ref{qoa2}), see, for example, \cite{BazHibKho02}. Therefore, the
$Q$-opera\-tors obtained from $L$-operators connected by the transformations (\ref{qoa1}),
(\ref{qoa2}) coincide.

Further, to construct $L$-operators of type $\hat L$ one can use homomorphisms of
two different six-pa\-ra\-me\-ter families,
\begin{gather}
\chi(h_{\delta - \alpha - \beta}) = - D_1 - D_2, \qquad \chi_\zeta(h_\alpha) = 2D_1 - D_2, \qquad
\chi_\zeta(h_\beta) = - D_1 + 2D_2,
\label{srcI1} \\[.5em]
\chi(e_{\delta - \alpha - \beta}) = \rho \, \mu_1 \, \mu_2 \, a_1 \, a_2 \, q^{(\nu_1 + \nu_2 -
1)D_1 +(\nu_2 + \nu_3 - 2) D_2},
\label{srcI2} \\[.5em]
\chi(e_\alpha) = \frac{1}{\mu_1} \, q^{- \nu_1 D_1 - \nu_2 D_2} a_1^\dagger, \qquad \chi(e_\beta) =
\frac{1}{\mu_2} \, q^{- (\nu_2 - 1)D_1 - \nu_3 D_2} a_2^\dagger, \label{srcI3}
\end{gather}
and
\begin{gather}
\chi(h_{\delta - \alpha - \beta}) = - D_1 - D_2, \qquad \chi_\zeta(h_\alpha) = 2D_1 - D_2, \qquad
\chi_\zeta(h_\beta) = - D_1 + 2 D_2,
\label{6} \\[.5em]
\chi(e_{\delta - \alpha - \beta})
= \rho \, \mu_1 \, \mu_2 \, a_1 \, a_2 \, q^{(\nu_1 + \nu_2 - 1)D_1 + (\nu_2 + \nu_3) D_2}, \\[.5em]
\chi(e_\alpha) = \frac{1}{\mu_1} \, q^{- \nu_1 D_1 - \nu_2 D_2} a_1^\dagger, \qquad \chi(e_\beta) =
\frac{1}{\mu_2} \, q^{- (\nu_2 + 1) D_1 - \nu_3 D_2} a_2^\dagger. \label{7}
\end{gather}
One can verify that these definitions are consistent with the defining relations of the algebra
$U_\hbar(\gothb'_+(A_2^{(1)}))$. One can freely change the parameters $\mu_1$, $\mu_2$ and $\nu_1$,
$\nu_2$, $\nu_3$ in the final expression by applying automorphisms (\ref{qoa1}), (\ref{qoa2}). As a
result we obtain equivalent $L$-operators. Changing the parameter $\rho$, we change the coefficient
at $\zeta^s$ in the final expression.

In order to construct $L$-operators of type $\check L$ we can use the
homomorphisms of two different six-pa\-ra\-me\-ter families,
\begin{gather}
\psi(h_{\delta - \alpha - \beta}) = - D_1 - D_2, \qquad \chi_\zeta(h_\alpha) = 2D_1 - D_2, \qquad
\chi_\zeta(h_\beta) = - D_1 + 2 D_2,
\label{8} \\[.5em]
\psi(f_{\delta - \alpha - \beta}) = \frac{\rho}{\mu_1 \mu_2} \, q^{- (\nu_1 + \nu_2 + 1)D_1 -(\nu_2
+ \nu_3 + 2) D_2} a_1^\dagger \, a_2^\dagger,
\\[.5em]
\psi(f_\alpha) = \mu_1 \, a_1 \, q^{\nu_1 D_1 + \nu_2 D_2}, \qquad \psi(f_\beta) = \mu_2 \, a_2 \,
q^{(\nu_2 + 1)D_1 + \nu_3 D_2}, \label{9}
\end{gather}
and
\begin{gather}
\psi(h_{\delta - \alpha - \beta}) = - D_1 - D_2, \qquad \chi_\zeta(h_\alpha) = 2D_1 - D_2, \qquad
\chi_\zeta(h_\beta) = - D_1 + 2D_2,
\label{17} \\[.5em]
\psi(f_{\delta - \alpha - \beta}) = \frac{\rho}{\mu_1 \mu_2} \, q^{- (\nu_1 + \nu_2 + 1)D_1 -(\nu_2
+ \nu_3) D_2} a_1^\dagger \, a_2^\dagger,
\\[.5em]
\psi(f_\alpha) = \mu_1 \, a_1 \, q^{\nu_1 D_1 + \nu_2 D_2}, \qquad \psi(f_\beta) = \mu_2 \, a_2 \,
q^{(\nu_2 - 1) D_1 + \nu_3 D_2}. \label{18}
\end{gather}
One can convince oneself that these definitions are consistent with the defining relations of the
algebra $U_\hbar(\gothb'_-(A_2^{(1)}))$. The parameters $\mu_1$, $\mu_2$ and $\nu_1$, $\nu_2$,
$\nu_3$ in the final expression can be freely changed by applying the automorphisms (\ref{qoa1}),
(\ref{qoa2}). Changing the parameter $\rho$, we change the coefficient at $\zeta^s$ in the final
expression.

As for the case of the quantum group $U_\hbar(\gothg'(A_1^{(1)}))$, there are a few methods to
obtain $L$-operators which are not equivalent to those which can be obtained by using the
homomorphisms defined above.

The automorphism group of the Dynkin diagram of the generalized Cartan matrix $A_2^{(1)}$ is
isomorphic to the dihedral group $\mathrm D_3$. This group coincides with the symmetric group
$\mathrm S_3$. Each automorphism $s \in \mathrm D_3$ gives rise to an automorphism of the quantum
group $U_\hbar(\gothg'(A_2^{(1)}))$ defined by the relations\footnote{We assume that the group
$\mathrm D_3$ is realized as a transformation group of the set $\{0, 1, 2\}$.}
\[
\sigma_s(h_{\alpha_i}) = h_{\alpha_{s(i)}}, \qquad \sigma_s(e_{\alpha_i}) = e_{\alpha_{s(i)}},
\qquad \sigma_s(f_{\alpha_i}) = f_{\alpha_{s(i)}}.
\]
Applying first one of the automorphisms $\sigma_s$, we modify the homomorphisms $\chi$ and $\psi$
defined above and obtain nonequivalent $L$-operators.

As before, if $\hat L(\zeta)$ is an $L$-operator of type $\hat L$, then the $L$-operator $\hat
L^{-1}(\zeta^{-1})$ is of type $\check L$, and if $\check L(\zeta)$ is an $L$-operator of type
$\check L$, then $\check L^{-1}(\zeta^{-1})$ is an $L$-operator of type $\hat L$.

At last, one can use the ani-involution $\tau \otimes \tau$ of the algebra $\Osc_\hbar
\otimes \Osc_\hbar$, where the anti-involution $\tau$ of the algebra $\Osc_\hbar$ is defined by the
relations (\ref{oaa}). Explicitly, one has
\[
\tau(a_i^{}) = a_i^\dagger, \qquad \tau(a_i^\dagger) = a_i^{}, \qquad \tau(D_i) = D_i, \qquad i = 1, 2.
\]
It is easy to determine that if $\hat L(\zeta)$ is an $L$-operator of type $\hat L$, then
$\tau(\hat L(\zeta^{-1}))$ is an $L$-operator of type $\check L$, and vice versa, if $\check
L(\zeta)$ is an $L$-operator of type $\check L$, then $\tau(\check L(\zeta^{-1}))$ is an
$L$-operator of type $\hat L$.

\subsubsection{Type $\hat L$}

We define a homomorphism $\chi$ from $U_\hbar(\gothb'_+(A_2^{(1)}))$ to $\Osc_\hbar \otimes
\Osc_\hbar$ using (\ref{srcI1})--(\ref{srcI3}) with $\mu_1 = \mu_2 = q - q^{-1}$, $\rho = 1 / (q -
q^{-1})^3$ and $\nu_1 = \nu_2 = \nu_3 = 0$. Then we use the relations (\ref{phiz1}) and
(\ref{phiz2}) to define homomorphisms $\chi_\zeta$. After all we have
\begin{gather}
\chi_\zeta(h_{\delta - \alpha - \beta}) = - D_1 - D_2, \qquad \chi_\zeta(h_\alpha) = 2D_1 - D_2,
\qquad \chi_\zeta(h_\beta) = - D_1 + 2D_2,
\label{cha2} \\
\chi_\zeta(e_{\delta - \alpha - \beta}) = \frac{1}{q - q^{-1}} \, a_1 \, a_2  \, q^{- D_1 - 2 D_2}
\, \zeta^{s - s_1 - s_2},
\label{ced2} \\
\chi_\zeta(e_\alpha) = \frac{1}{q - q^{-1}} \, a_1^\dagger \, \zeta^{s_1}, \qquad
\chi_\zeta(e_\beta) = \frac{1}{q - q^{-1}} \, q^{D_1} a_2^\dagger \, \zeta^{s_2}. \label{cea2}
\end{gather}

First, we find the expression for $\chi_{\zeta_1} \otimes \varphi_{\zeta_2} (\calK)$. Taking into
account equations (\ref{k2}), (\ref{phih2}) and (\ref{cha2}), we obtain
\begin{equation}
\chi_{\zeta_1} \otimes \varphi_{\zeta_2} (\calK) = q^{D_1} E_{11} + q^{- D_1 + D_2} E_{22} + q^{-D_2} E_{33}.
\label{ko}
\end{equation}

The definition (\ref{eab}), with an account of the relations (\ref{cea2}), gives
\[
\chi_\zeta(e_{\alpha + \beta}) = 0,
\]
and using (\ref{eda2}) we obtain
\begin{gather}
\chi_\zeta(e_{\delta - \alpha}) = \frac{1}{q - q^{-1}} \, q^{-2} a_1 \, q^{-2 D_2}
\zeta^{s - s_1},
\label{1a} \\
\chi_\zeta(e_{\delta - \beta}) = \frac{1}{q - q^{-1}} \, q^{-1} a_2 \, q^{- D_1 - 2 D_2} \zeta^{s - s_2}.
\label{1b}
\end{gather}
Now, the definition (\ref{e3}) gives
\begin{gather*}
\chi_\zeta(e'_{\delta, \, \alpha}) = \frac{1}{q - q^{-1}} \, q^{-3} q^{-2 D_2} \zeta^{s}, \qquad
\chi_\zeta(e'_{\delta, \, \beta})
= \frac{1}{q - q^{-1}} \, q^{-4} [(1 + q^2) q^{-2 D_2} - q^2] \, \zeta^{s}, \\[.5em]
\chi_\zeta(e'_{\delta, \, \alpha + \beta}) = 0,
\end{gather*}
and using the definitions (\ref{e4}) and (\ref{e5}) we determine that for $m > 0$ we have
\begin{gather*}
\chi_\zeta(e_{\beta + m \delta})
= \frac{1}{q - q^{-1}} \, q^{- 4 m} a_2^\dagger \, q^{D_1 - 2 m D_2} \zeta^{s_2 + ms}, \\[.5em]
\chi_\zeta(e_{\alpha + m \delta}) = 0, \qquad \chi_\zeta(e_{\alpha + \beta + m \delta}) = 0, \\[.5em]
\chi_\zeta(e_{(\delta - \beta) + m \delta})
= \frac{1}{q - q^{-1}} \, q^{- 2 m - 1} a_2 \, q^{-D_1 - 2 (m + 1) D_2} \zeta^{(s - s_2) + m s}, \\[.5em]
\chi_\zeta(e_{(\delta - \alpha) + m\delta}) = 0,
\qquad \chi_\zeta(e_{(\delta - \alpha - \beta) + m \delta}) = 0.
\end{gather*}
Now, taking into account the relations (\ref{phizf21})--(\ref{phizf22}), we come to the equations
\begin{gather}
\chi_{\zeta_1} \otimes \varphi_{\zeta_2} (\calR_{\alpha, \, 0}) = 1 + a_1^\dagger \,
\zeta^{s_1}_{12} \, E_{21},
\label{chir1} \\[.5em]
\chi_{\zeta_1} \otimes \varphi_{\zeta_2} (\calR_{\alpha, \, m}) = 1, \qquad m > 0, \\[.5em]
\chi_{\zeta_1} \otimes \varphi_{\zeta_2} (\calR_{\beta, \, m})
= 1 + q^{- 2 m} a_2^\dagger \, q^{D_1 -  2 m D_2} \, \zeta^{s_2 + ms}_{12} E_{32}, \\[.5em]
\chi_{\zeta_1} \otimes \varphi_{\zeta_2} (\calR_{\alpha + \beta, \, m}) = 1, \\[.5em]
\chi_{\zeta_1} \otimes \varphi_{\zeta_2} (\calR_{\delta - \alpha, \, 0})
= 1 + q^{-2} a_1 q^{-2 D_2} \zeta^{s - s_1}_{12} E_{12}, \\[.5em]
\chi_{\zeta_1} \otimes \varphi_{\zeta_2} (\calR_{\delta - \alpha, \, m}) = 1, \qquad m > 0, \\[.5em]
\chi_{\zeta_1} \otimes \varphi_{\zeta_2} (\calR_{\delta - \beta, \, m})
= 1 - a_2 \, q^{-D_1 - 2 (m + 1) D_2} \, \zeta^{(s - s_2) + m s}_{12} E_{23}, \\[.5em]
\chi_{\zeta_1} \otimes \varphi_{\zeta_2} (\calR_{\delta - \alpha - \beta, \, 0})
= 1 + a_1 a_2 \, q^{- D_1 - 2 D_2} \, \zeta^{s - s_1 - s_2}_{12} E_{13}, \\[.5em]
\chi_{\zeta_1} \otimes \varphi_{\zeta_2} (\calR_{\delta - \alpha - \beta, \, m}) = 1, \qquad m > 0.
\label{chir2}
\end{gather}
One can determine that, due to the defining properties of $E_{ab}$, $\chi_{\zeta_1} \otimes
\varphi_{\zeta_2}(\calR_{\alpha, \, m})$ and $\chi_{\zeta_1} \otimes
\varphi_{\zeta_2}(\calR_{\alpha + \beta, \, n})$ commute for any $m$ and $n$. Therefore, we can
rearrange the factors entering $\chi_{\zeta_1} \otimes \varphi_{\zeta_2}(\calR_{\prec \delta})$ in
such a way that the factors corresponding to the roots $\alpha + m \delta$ come first, then the
factors corresponding to the roots $\alpha + \beta + m \delta$, and finally the factors
corresponding to the roots $\beta + m \delta$. Similarly, we can rearrange the factors entering
$\chi_{\zeta_1} \otimes \varphi_{\zeta_2}(\calR_{\succ \delta})$ in such a way that the factors
corresponding to the roots $(\delta - \beta) + m \delta$ come first, then the factors corresponding
to the roots $(\delta - \alpha) + m \delta$ and finally the factors corresponding to the roots
$(\delta - \alpha - \beta) + m \delta$.

We denote by $\hat L_\gamma(\zeta_{12})$, $\gamma \in \Delta_+(A_2)$, the product of the factors
corresponding to the roots $\gamma + m \delta$, and by $\hat L_{\delta - \gamma}(\zeta_{12})$,
$\gamma \in \Delta_+(A_2)$, the product of the factors corresponding to the roots $(\delta -
\gamma) + m \delta$. Now, using (\ref{chir1})--(\ref{chir2}) and summing up the arising geometrical
series, we come to the expressions
\begin{gather}
\hat L_\alpha(\zeta) = 1 + a_1^\dagger \, \zeta^{s_1} E_{21},
\label{3} \\[.5em]
\hat L_\beta(\zeta) = 1 + a_2^\dagger \, q^{D_1} (1 - q^{-2} q^{- 2 D_2} \, \zeta^s)^{-1}
 \zeta^{s_2}\, E_{32}, \\[.5em]
\hat L_{\alpha + \beta}(\zeta) = 1, \\[.5em]
\hat L_{\delta - \alpha}(\zeta)
= 1 + q^{-2} a_1 \, q^{-2 D_2} \, \zeta^{s - s_1} E_{12}, \\[.5em]
\hat L_{\delta - \beta}(\zeta)
= 1 - a_2 \, q^{- D_1 - 2 D_2} (1 - q^{-2D_2} \, \zeta^s)^{-1} \zeta^{s - s_2} E_{23}, \\[.5em]
\hat L_{\delta - \alpha - \beta}(\zeta) = 1 + a_1 a_2 \, q^{- D_1 - 2 D_2} \, \zeta^{s - s_1 - s_2} E_{13}.
\label{4}
\end{gather}

Recalling the definition (\ref{e6}) and having in mind equations (\ref{1a}) and (\ref{1b}), we
determine that for $m > 1$ we have
\begin{gather*}
\chi_\zeta(e'_{m \delta, \, \alpha}) = 0, \\
\chi_\zeta(e'_{m \delta, \, \beta}) = - \frac{1}{(q - q^{-1})^2} \, q^{-4 m - 1} [(1 - q^{2(m +
1)}) q^{- 2 m D_2} - q^2 (1 - q^{2m}) q^{-2(m - 1) D_2} ] \zeta^{m s}.
\end{gather*}
Starting with these expressions we obtain
\begin{gather*}
\chi_\zeta(e_{m \delta, \, \alpha})
= \frac{1}{q - q^{-1}} \, (-1)^{m - 1} q^{- 3 m} q^{-2 m D_2} \frac{\zeta^{ms}}{m},\\
\chi_\zeta(e_{m \delta, \, \beta})
= \frac{1}{q - q^{-1}} \, q^{-2 m} [(1 + q^{- 2 m}) q^{-2mD_2} - 1] \frac{\zeta^{ms}}{m}.
\end{gather*}
Denoting $\hat L_\delta(\zeta_{12}) = \chi_{\zeta_1} \otimes \varphi_{\zeta_2}(\calR_{\sim
\delta})$, we come to the equation
\begin{equation}
\hat L_\delta(\zeta) = \rme^{\lambda_3(q^{-2} \zeta^s)} [E_{11} + (1 - q^{- 2}
q^{-2D_2} \zeta^s) E_{22} + (1 - \zeta^s) (1 - q^{-2D_2} \zeta^s)^{-1} E_{33}]. \label{5}
\end{equation}
Multiplying the factors (\ref{3})--(\ref{4}), (\ref{5}) and (\ref{ko}) in the given order we
obtain the following $L$-operator
\[ \hat L(\zeta) = \rme^{\lambda_3(q^{-2} \zeta^s)} \left( \begin{array}{ccc} q^{D_1} &
q^{-2} a_1 \, q^{- D_1 - D_2} \, \zeta^{s - s_1}
& a_1 a_2 \, q^{- D_1 - 3 D_2} \, \zeta^{s - s_1 - s_2} \\[.5em]
a_1^\dagger \, q^{D_1} \, \zeta^{s_1} & q^{- D_1 + D_2} - q^{-2} \, q^{D_1 - D_2} \zeta^s  & - a_2
\, q^{D_1 - 3 D_2} \, \zeta^{s - s_2} \\[.5em]
0 & a_2^\dagger \, q^{D_2} \, \zeta^{s_2} & q^{-D_2}
\end{array} \right).
\]
As we noted above its inverse after the change $\zeta \to \zeta^{-1}$ becomes an $L$-operator of
type $\check L$. To write it in a simple form we apply the automorphism
\[
a_i \to q^{-1} a_i \, q^{2 D_i}, \qquad a^\dagger_i \to q \, q^{-2 D_i} \, a^\dagger_i, \qquad i =
1,2,
\]
being a particular case of the automorphisms (\ref{qoa1}), (\ref{qoa2}). After all we come to the
expression
\[ \check L(\zeta) = \frac{\rme^{-\lambda_3(q^{-2} \zeta^{-s})}}{1 - \zeta^s}
\left( \begin{array}{ccc} q^2 q^{D_1} - q^{-D_1} \, \zeta^s & a_1 \, q^{D_1} \, \zeta^{s_1}
& q^{-1} a_1 a_2 \, \zeta^{s_1 + s_2} \\[.5em]
a_1^\dagger \, q^{- D_1 - D_2} \, \zeta^{s - s_1} & - q^{D_1 - D_2} \zeta^s & - a_2
\, q^{- D_2} \, \zeta^{s_2} \\[.5em]
- a^\dagger_1 a^\dagger_2 \, q^{- D_1 - D_2} \, \zeta^{s - s_1 - s_2} & a_2^\dagger \, q^{D_1 -
D_2} \, \zeta^{s - s_2} & q^{-D_2} - q^{D_2} \, \zeta^s
\end{array} \right).
\]

One can verify that
\begin{equation}
\hat L^{(s, s_1, s_2)}(\zeta_{12}) = \gamma_{\zeta_1}(G(\zeta_2) \hat L^{(1, 0, 0)}(\zeta_{12}^s)
G^{-1}(\zeta_2)), \label{hlsss}
\end{equation}
where the matrix $G(\zeta)$ is given by equation (\ref{gz2}), and the mapping $\gamma_\zeta$,
$\zeta \in \bbC^\times$, is defined as
\begin{equation}
\gamma_\zeta(a_i) = a_i \zeta^{-s_i}, \qquad \gamma_\zeta(a^\dagger_i) = a^\dagger_i \zeta^{s_i},
\qquad \gamma_\zeta(D_i) = D_i, \qquad i = 1, 2. \label{sgz2}
\end{equation}
Therefore, the $Q$-operators obtained from $L$-operators corresponding to different values of $s$,
$s_1$ and $s_2$ are connected by a change of the spectral parameters and a similarity
transformation.

If we use the family (\ref{6})--(\ref{7}) with $\mu_1 = \mu_2 = q - q^{-1}$, $\rho = - 1 / (q -
q^{-1})^3$ and $\nu_1 = \nu_2 = \nu_3 = 0$ we come to homomorphisms $\chi_\zeta$ defined in the
following way:
\begin{gather*}
\chi_\zeta(h_{\delta - \alpha - \beta}) = - D_1 - D_2, \qquad \chi_\zeta(h_\alpha) = 2D_1 - D_2,
\qquad \chi_\zeta(h_\beta) = - D_1 + 2D_2, \\
\chi_\zeta(e_{\delta - \alpha - \beta})
= - \frac{1}{q - q^{-1}} \, a_1 a_2 \, q^{- D_1} \zeta^{s - s_1 - s_2}, \\
\chi_\zeta(e_\alpha) = \frac{1}{q - q^{-1}} \, a_1^\dagger \, \zeta^{s_1}, \qquad
\chi_\zeta(e_\beta) = \frac{1}{q - q^{-1}} \, q^{- D_1} a_2^\dagger \, \zeta^{s_2}.
\end{gather*}
This leads to the $L$-operator
\[
\hat L(\zeta) = \frac{\rme^{- \lambda_3(q^2 \zeta^s)}}{1 - \zeta^s} \left( \begin{array}{ccc}
q^{D_1} - q^{-2} q^{-D_1} \zeta^s & - a_1 \, q^{- 3 D_1 + D_2} \,
\zeta^{s - s_1} & - a_1 a_2 \, q^{- D_1 - D_2} \, \zeta^{s - s_1 - s_2} \\[.5em]
a_1^\dagger \, q^{D_1} \, \zeta^{s_1} & q^{- D_1 + D_2} & a_2 \, q^{D_1 - D_2} \,
\zeta^{s - s_2} \\[.5em]
q^{-1} a_1^\dagger a_2^\dagger \, \zeta^{s_1 + s_2} & a_2^\dagger \, q^{- 2 D_1 + D_2} \,
\zeta^{s_2} & q^{-D_2} - q^{D_2} \zeta^s
\end{array} \right)
\]
which also satisfies equation (\ref{hlsss}). Besides, with the change $\zeta \to \zeta^{-1}$ it can
be verified that the corresponding inverse $L$-operator satisfies the relation (\ref{crll}) of
$\check{L}$-type.

\subsubsection{Type $\check L$}

We define a homomorphism $\psi$ from $U_\hbar(\gothb'_-(A_2^{(1)}))$ to $\Osc_\hbar \otimes
\Osc_\hbar$ using (\ref{8})--(\ref{9}) with $\mu_1 = \mu_2 = 1 / (q - q^{-1})$, $\rho = 1 / (q -
q^{-1})^3$ and $\nu_1 = \nu_2 = \nu_3 = 0$. For the corresponding homomorphisms $\psi_\zeta$ we
then have
\begin{gather}
\psi_\zeta(h_{\delta - \alpha - \beta}) = - D_1 - D_2, \qquad \psi_\zeta(h_\alpha) = 2 D_1 - D_2,
\qquad \psi_\zeta(h_\beta) = - D_1 + 2 D_2,
\label{10} \\
\psi_\zeta(f_{\delta - \alpha - \beta})
= \frac{1}{q - q^{-1}} \, q^{- D_1 - 2 D_2} a_1^\dagger a_2^\dagger \, \zeta^{- s + s_1 + s_2}, \\
\psi_\zeta(f_\alpha) = \frac{1}{q - q^{-1}} \, a_1 \zeta^{-s_1}, \qquad \psi_\zeta(f_\beta)
= \frac{1}{q - q^{-1}} \, a_2 \, q^{D_1} \zeta^{-s_2}.
\label{11}
\end{gather}

The expression for $\varphi_{\zeta_1} \otimes \psi_{\zeta_2}(\calK)$ is the same as in the previous
case and has the form (\ref{ko}).

It follows from (\ref{fab}) that
\[
f_{\alpha + \beta} = f_\beta \, f_\alpha - q \, f_\alpha \, f_\beta,
\]
hence, taking into account equations (\ref{11}), we obtain
\[
\psi_\zeta(f_{\alpha + \beta}) = - \frac{1}{q - q^{-1}} \, a_1 \, a_2 \, q^{D_1} \zeta^{- s_1 - s_2}.
\]
Equation (\ref{fdg}) leads to the relations
\[
f_{\delta - \alpha}
= f_{\delta - \alpha - \beta} \, f_\beta - q \, f_\beta \, f_{\delta - \alpha - \beta},
\qquad f_{\delta - \beta}
= f_{\delta - \alpha - \beta} \, f_\alpha - q \, f_\alpha \, f_{\delta - \alpha - \beta}
\]
which give
\begin{equation}
\psi_\zeta(f_{\delta - \alpha}) = \frac{1}{q - q^{-1}} \, a_1^\dagger \, \zeta^{-s + s_1}, \qquad
\psi_\zeta(f_{\delta - \beta}) = \frac{1}{q - q^{-1}} \, q^{-1} a_2^\dagger \, q^{D_1 - 2 D_2} \,
\zeta^{-s + s_2}. \label{13}
\end{equation}
Using equation (\ref{f6}), we come to the relations
\begin{gather*}
\psi_\zeta(f'_{\delta, \, \alpha}) = \frac{1}{q - q^{-1}} \, q [(1 + q^2) q^{2 D_1} - 1]
\zeta^{-s},
\qquad \psi_\zeta(f'_{\delta, \, \beta}) = \frac{1}{q - q^{-1}} \, q^2 q^{2 D_1} \zeta^{-s}, \\
\psi_\zeta(f'_{\delta, \, \alpha + \beta}) = \frac{1}{q - q^{-1}} q [(1 + q^2) q^{2 D_1} - q^{2 D_1
- 2 D_2} - 1] \zeta^{-s},
\end{gather*}
and, using (\ref{f4}) and (\ref{f5}), we determine that for $m > 0$ we have
\begin{gather*}
\psi_\zeta(f_{\alpha + m \delta}) = \frac{1}{q - q^{-1}} (-1)^m q^m a_1 q^{2 m D_1} \zeta^{- s_1 - m s}, \\
\psi_\zeta(f_{\alpha + \beta + m \delta}) = \frac{1}{q - q^{-1}} (-1)^{m - 1} q^m a_1 a_2 q^{(2 m + 1)D_1}
\zeta^{- s_1 - s_2 - m s}, \\
\psi_\zeta(f_{\beta + m \delta}) = 0, \\[.5em]
\psi_\zeta(f_{(\delta - \beta) + m \delta}) = 0, \\
\psi_\zeta(f_{(\delta - \alpha) + m \delta})
= \frac{1}{q - q^{-1}} (-1)^m q^{3 m} a_1^\dagger q^{2 m D_1} \zeta^{- (s - s_1) - m s}, \\
\psi_\zeta(f_{(\delta - \alpha - \beta) + m \delta})
= \frac{1}{q - q^{-1}} (-1)^m q^{3 m - 3} a_1^\dagger a_2^\dagger q^{(2 m - 1)D_1 - 2 D_2}
\zeta^{- (s - s_1 - s_2) - m s}.
\end{gather*}
Taking into account the relations (\ref{phizeamd2})--(\ref{12}), we come to the equations
\begin{gather*}
\varphi_{\zeta_1} \otimes \psi_{\zeta_2} (\calR_{\alpha, \, m})
= 1 + a_1 q^{2 m D_1} \zeta^{s_1 + ms}_{12} E_{12}, \\[.5em]
\varphi_{\zeta_1} \otimes \psi_{\zeta_2} (\calR_{\alpha + \beta, \, m})
= 1 -  a_1 a_2 \, q^{(2 m + 1)D_1} \zeta^{s_1 + s_2 + m s}_{12} E_{13}, \\[.5em]
\varphi_{\zeta_1} \otimes \psi_{\zeta_2} (\calR_{\beta, \, 0})
= 1 + a_2 \, q^{D_1} \zeta^{s_2 + ms}_{12} E_{23}, \\[.5em]
\varphi_{\zeta_1} \otimes \psi_{\zeta_2} (\calR_{\beta, \, 0}) = 1, \qquad m > 0, \\[.5em]
\varphi_{\zeta_1} \otimes \psi_{\zeta_2} (\calR_{\delta - \beta, \, 0})
= 1 - q^{- 2} a_2^\dagger q^{D_1 - 2 D_2} \zeta^{s - s_2}_{12} E_{32}, \\[.5em]
\varphi_{\zeta_1} \otimes \psi_{\zeta_2} (\calR_{\delta - \beta, \, m}) = 1, \qquad m > 0, \\[.5em]
\varphi_{\zeta_1} \otimes \psi_{\zeta_2} (\calR_{\delta - \alpha, \, m})
= 1 + q^{2 m} a_1^\dagger q^{2 m D_1} \zeta^{(s - s_1) + m s} E_{21}, \\[.5em]
\varphi_{\zeta_1} \otimes \psi_{\zeta_2} (\calR_{\delta - \alpha - \beta, \, m})
= 1 + q^{2 m - 3} a_1^\dagger a_2^\dagger \, q^{(2 m - 1)D_1 - 2 D_2} \zeta^{(s - s_1 - s_2) + m s} E_{31}.
\end{gather*}
Now we use the appropriate reordering of the factors entering $\varphi_{\zeta_1} \otimes
\psi_{\zeta_2}(\calR_{\prec \delta})$ and $\varphi_{\zeta_1} \otimes \psi_{\zeta_2}(\calR_{\succ
\delta})$ to determine that
\begin{gather*}
\varphi_{\zeta_1} \otimes \psi_{\zeta_2}(\calR_{\prec \delta})
= \check L_\alpha(\zeta_{12}) \check L_{\alpha + \beta}(\zeta_{12}) \check L_\beta(\zeta_{12}), \\
\varphi_{\zeta_1} \otimes \psi_{\zeta_2}(\calR_{\succ \delta}) = \check
L_{\delta - \beta}(\zeta_{12}) \check L_{\delta - \alpha}(\zeta_{12}) \check L_{\delta - \alpha -
\beta}(\zeta_{12}),
\end{gather*}
where
\begin{gather}
\check L_\alpha(\zeta) = 1 + a_1 \zeta^{s_1} (1 - q^{2 D_1} \zeta^s)^{-1} E_{12},
\label{14} \\[.5em]
\check L_{\alpha + \beta}(\zeta)
= 1 - a_1 a_2 \, q^{D_1} \zeta^{s_1 + s_2} (1 - q^{2 D_1} \zeta^s)^{-1} E_{13}, \\[.5em]
\check L_\beta(\zeta) = 1 + a_2 \, q^{D_1} \zeta^{s_2} E_{23}, \\[.5em]
\check L_{\delta - \beta}(\zeta) = 1 - q^{-2} a_2^\dagger \, q^{D_1 - 2 D_2} \zeta^{s - s_2} E_{32}, \\[.5em]
\check L_{\delta - \alpha} = 1 + a_1^\dagger \zeta^{s - s_1} (1 - q^2 q^{2 D_1} \zeta^s)^{-1} E_{21}, \\[.5em]
\check L_{\delta - \alpha - \beta} = 1 + q^{-3} a_1^\dagger a_2^\dagger \, q^{- D_1 - 2 D_2}
\zeta^{s - s_1 - s_2} (1 - q^2 q^{2 D_1} \zeta^s)^{-1}  E_{31}. \label{15}
\end{gather}

Using equation (\ref{f6}) and taking into account (\ref{13}), we determine that for $m > 1$ one has
\begin{gather*}
\psi_\zeta(f'_{m \delta, \, \alpha}) = \frac{1}{(q - q^{-1})^2} \,
(-1)^m q^{m - 1} [(1 - q^{2(m + 1)}) q^{2 m D_1} - (1 - q^{2m}) q^{2(m - 1) D_1} ] \zeta^{-m s}, \\
\psi_\zeta(f'_{m \delta, \, \beta}) = 0.
\end{gather*}
Having in mind the relation (\ref{f7}), we obtain
\begin{gather*}
\psi_\zeta(f_{m \delta, \, \alpha})
= \frac{1}{q - q^{-1}} \, (-1)^{m - 1} q^m [(1 + q^{2 m}) q^{2 m D_1} - 1] \frac{\zeta^{-m s}}{m}, \\
\psi_\zeta(f_{m \delta, \, \beta})
= \frac{1}{q - q^{-1}} \, q^{2 m} q^{2 m D_1} \frac{\zeta^{-m s}}{m}.
\end{gather*}
Now, denoting $\check L_\delta(\zeta_{12}) = \varphi_{\zeta_1} \otimes \psi_{\zeta_2}(\calR_{\sim
\delta})$, we come to the equation
\begin{equation}
\check L_\delta(\zeta) = \rme^{\lambda_3(q^{-2} \zeta^s)} [(1 - \zeta^s)(1 - q^2 q^{2 D_1}
\zeta^s)^{-1} E_{11} + (1 - q^{ 2D_1} \zeta^s) E_{22} + E_{33}]. \label{16}
\end{equation}
Multiplying the factors (\ref{14})--(\ref{15}), (\ref{16}) and (\ref{ko}) in the
prescribed order we obtain the following $L$-operator
\[
\check L(\zeta) = \rme^{\lambda_3(q^{-2} \zeta^s)} \left( \begin{array}{ccc}
q^{D_1} & a_1 \, q^{- D_1 + D_2} \, \zeta^{s_1} & 0 \\[.5em]
q^{-2} \, a_1^\dagger \, q^{D_1 - 2 D_2} \, \zeta^{s - s_1}
& q^{- D_1 + D_2} - q^{-2} q^{D_1 - D_2} \zeta^s & a_2 \, q^{D_1 - D_2} \, \zeta^{s_2} \\[.5em]
q^{-3} \, a_1^\dagger a_2^\dagger \, q^{- 2 D_2} \,
\zeta^{s - s_1 - s_2} & - q^{-2} \, a_2^\dagger \, q^{- D_2} \, \zeta^{s - s_2} & q^{-D_2}
\end{array} \right).
\]
In this case we have
\begin{equation}
\check L^{(s, s_1, s_2)}(\zeta_{12})
= G(\zeta_1) \gamma_{\zeta_2}(\check L^{(1, 0, 0)}(\zeta_{12}^s)) G^{-1}(\zeta_1),
\label{clsss}
\end{equation}
where the matrix $G(\zeta)$ is given by equation (\ref{gz2}) and the mapping $\gamma_\zeta$ by the
relations (\ref{sgz2}).

If we use (\ref{17})--(\ref{18}) with $\mu_1 = \mu_2 = q - q^{-1}$, $\rho = - 1 / (q - q^{-1})^3$
and $\nu_1 = \nu_2 = \nu_3 = 0$ we come to homomorphisms $\psi_\zeta$ defined in the following way:
\begin{gather*}
\psi_\zeta(h_{\delta - \alpha - \beta}) = - D_1 - D_2, \qquad \psi_\zeta(h_\alpha)
= 2D_1 - D_2, \qquad \psi_\zeta(h_\beta) = - D_1 + 2D_2, \\
\psi_\zeta(f_{\delta - \alpha - \beta})
= - \frac{1}{q - q^{-1}} \, q^{- D_1} \, a_1^\dagger \, a_2^\dagger \, \zeta^{- s + s_1 + s_2}, \\
\psi_\zeta(f_\alpha) = \frac{1}{q - q^{-1}} \, a_1 \, \zeta^{-s_1},
\qquad \psi_\zeta(f_\beta) = \frac{1}{q - q^{-1}} \, a_2 \, q^{- D_1} \zeta^{-s_2},
\end{gather*}
and we obtain one more $L$-operator
\[
\check L(\zeta) = \frac{\rme^{- \lambda_3(q^2 \zeta^s)}}{1 - \zeta^s}
\left( \begin{array}{ccc}
q^{D_1} - q^{-2} q^{-D_1} \zeta^s & a_1 \, q^{- D_1 + D_2} \,
\zeta^{s_1} & a_1 a_2 \, q^{- D_1 - D_2} \, \zeta^{s_1 + s_2} \\[.5em]
- q^{-2} \, a_1^\dagger \, q^{-D_1} \,
\zeta^{s - s_1} & q^{- D_1 + D_2} & a_2 \, q^{- D_1 - D_2} \, \zeta^{s_2} \\[.5em]
- q^{-1} \, a_1^\dagger a_2^\dagger \, \zeta^{s - s_1 - s_2} & a_2^\dagger \, q^{D_2}
\, \zeta^{s - s_2} & q^{-D_2} - q^{D_2} \zeta^s
\end{array} \right),
\]
which also satisfies equation (\ref{clsss}).

\section{\texorpdfstring{Concluding remarks}{Concluding remarks}}
\label{TheEnd}

Note here that, as is known, the $R$-matrices considered in this paper, up to a respective scalar
factor and with a special choice of the parameters $s_i$, allow for the decomposition
\begin{equation}
R(\zeta) = \zeta \, R_0 - \zeta^{-1} R_0^{-1}, \label{c1}
\end{equation}
where the non-degenerate matrix $R_0$ does not depend on the spectral parameter and has the form
\begin{equation}
R_0 = \sum^n_{a, b = 1} q^{\delta_{a b}} E_{aa} \otimes E_{bb} + (q - q^{-1})
\sum_{a < b} E_{ab} \otimes E_{ba}. \label{c2}
\end{equation}
For the cases considered in our paper $n$ is either $2$ or $3$, but actually the relations
(\ref{c1}) and (\ref{c2}) are valid for all quantum groups $U_\hbar(\gothg(A_{n-1}^{(1)}))$.
Obviously, the same decomposition holds also for $\hat R(\zeta)$  with
\[
\hat{R}_0 = \sum^n_{a, b = 1} q^{\delta_{ab}} E_{ab} \otimes E_{ba}
+ (q - q^{-1}) \sum_{a < b} E_{aa} \otimes E_{bb}.
\]
The $L$-operators have similar property, and, for example, an $L$-operator of type $\hat L$, again
up to a respective scalar factor, can be represented in the form
\[
\hat L( \zeta) = \zeta \, \hat L_+ - \zeta^{-1} \hat L_-,
\]
where $\hat L_+$ and $\hat L_-$ do not depend on the spectral parameter and satisfy the relations
\begin{align*}
& \hat R_0 (\hat L_+ \times \hat L_+) = (\hat L_+ \times \hat L_+) \hat R_0, \\
& \hat R_0 (\hat L_- \times \hat L_-) = (\hat L_- \times \hat L_-) \hat R_0, \\
& \hat R_0 (\hat L_+ \times \hat L_-) = (\hat L_- \times \hat L_+) \hat R_0.
\end{align*}
where $\times$ means a generalization of the Kronecker product defined by equation (\ref{akp}). One
could try to construct $L$-operators starting from these relations. In this way, however, one would
lose the convenient and highly non-trivial normalization which is implied by the functional
relations in the universal form arising when they are derived from the universal $R$-matrix.

Moreover, it is worthwhile noticing that the matrices $\hat L_+$ and $\hat L_-$ have
upper-triangu\-lar and lower-triangular forms, respectively. In the considered $q$-oscillator
representation the matrix $\hat L_+$ turns out to be degenerate and the matrix $\hat L_-$ is
non-degenerate. Moreover, the matrix $\hat \Pi = \hat L_-^{-1} \hat L_+$ satisfies the relation
$\hat \Pi^2 = \hat \Pi$. This equation implies that the matrix $\hat L(1)$ is singular. It seems
that this property is important for the very existence of the functional equations satisfied by the
corresponding transfer matrices and $Q$-operators, see, for example, \cite{ProSer01}.

Note that for an $L$-operator of type $\check L$ on the other hand the matrix $\check L_-$ turns
out to be degenerate and the matrix $\check L_+$ is non-degenerate. Here the matrix $\check \Pi =
\check L_+^{-1} \check L_-$ satisfies the relation $\check \Pi^2 = \check \Pi$, and the matrix
$\check L(1)$ is singular as well.

\appendix

\section{Kac-Moody algebras of finite and affine type}

\subsection{Generators and relations}

Let $A = (a_{ij})$ be a generalized Cartan matrix of finite or affine type. It is customary to
assume that  the numeration of the indices of $A$ starts from $1$ for the finite type, and from $0$
for the affine type. Denote by $\gothg'(A)$ the complex Lie algebra defined by $3n$ generators
$h_i$, $e_i$, $f_i$ and by the relations
\begin{gather*}
[h_i, h_j] = 0, \\[.5em]
[h_i, e_j] = a_{ij} \, e_j, \qquad [h_i, f_j] = - a_{ij} \, f_j, \\[.5em]
[e_i, f_j] = \delta_{ij} \, h_i, \\[.5em]
(\ad \, e_i)^{1 - a_{ij}} (e_j) = 0, \qquad (\ad \, f_i)^{1 - a_{ij}} (f_j) = 0.
\end{gather*}
It is assumed that $i$ and $j$ are different in the last line of the relations called {\em Serre
relations\/}. For the affine type we assume that the last $n - 1$ rows of the matrix $A$ are
linearly independent which can be achieved by applying a permutation to the rows and the same
permutation to the columns of $A$. In this case we introduce an additional generator $d$ and
additional relations
\[
[d, e_i] = \delta_{0i} e_i, \qquad [d, f_i] = - \delta_{0i} f_i, \qquad [d, h_i] = 0.
\]
We denote the corresponding Lie algebra by $\gothg(A)$ and assume that $\gothg(A) = \gothg'(A)$ for
finite type. The Lie algebra $\gothg(A)$ is the Kac-Moody algebra with the generalized Cartan
matrix $A$. In the finite-type case $\gothg(A)$ is isomorphic to the corresponding
finite-dimensional complex simple Lie algebra with the Cartan matrix $A$.

\subsection{Cartan subalgebra and roots}

The linear span $\gothh(A)$ of the generators $h_i$ for the finite type, or of the generators $h_i$
and $d$ for the affine type, is called the {\em Cartan subalgebra\/}. We denote the dual space of
$\gothh(A)$ by $\gothh^*(A)$. It can be shown that
\[
\gothg(A) = \gothh(A) \oplus \bigoplus_{\gamma \in \Delta(A)} \gothg_\gamma(A),
\]
where for any $\gamma \in \gothh^*(A)$ we denote
\[
\gothg_\gamma(A) = \{ x \in \gothg(A) \mid [h, x] = \gamma(h) x \mbox{ for all } h \in \gothh(A) \}
\]
and
\[
\Delta(A) = \{ \gamma \in \gothh^*(A) \mid \gamma \ne 0, \, \gothg_\gamma \ne \{ 0 \} \}.
\]
The elements of $\Delta(A)$ are called {\em roots\/} of $\gothg(A)$. The subspace
$\gothg_\gamma(A)$ for $\gamma \in \Delta(A)$ is said to be the {\em root space\/} of $\gamma$, and
its nonzero elements are called {\em root vectors\/}.

It is clear that $e_i$ are root vectors. Denote the corresponding roots by $\alpha_i$. These roots
are called {\em simple\/}. Any other root is a linear combination of simple roots with integer
coefficients all of which are either non-negative or non-positive. In the former case we say that
the root is {\em positive\/} and in the latter that it is {\em negative\/}. In particular, $f_i$
are root vectors corresponding to the negative roots $-\alpha_i$. One can write $\Delta(A) =
\Delta_+(A) \sqcup \Delta_-(A)$, where $\Delta_+(A)$ and $\Delta_-(A)$ are the sets of positive and
negative roots respectively.

The set $Q(A) \subset \gothh^*(A)$, defined as
\[
Q(A) = \bigoplus_{i=1}^n \bbZ \, \alpha_i,
\]
is called the {\em root lattice\/}. We denote by $Q_+(A)$ and $Q_-(A)$ the sublattices
\[
Q_+(A) = \bigoplus_{i=1}^n \bbN \, \alpha_i, \qquad Q_-(A) = \bigoplus_{i=1}^n \bbN \, (-\alpha_i).
\]

\subsection{Bilinear form}
\label{a:A.3}

It is known that a generalized Cartan matrix of finite or affine type is symmetrizable. This means
that there exist unique co-prime positive integers $d_i$ such that the matrix $(d_i a_{ij})$ is
symmetric. One defines a non-degenerate symmetric bilinear form $(\cdot, \cdot)$ on the Cartan
subalgebra $\gothh(A)$ by the equation
\[
(h_i, h_j) = a_{ij}^{} \, d^{-1}_j
\]
for the finite-type case, and additionally by the equations
\[
(h_i, d) = \delta_{i0}^{} \, d^{-1}_0, \qquad (d, d) = 0
\]
for the affine-type case. In both cases, with respect to the corresponding bilinear form on
$\gothh^*(A)$, one has
\[
(\alpha_i, \alpha_j) = d_i a_{ij}.
\]

\subsection{Extended Cartan matrix}
\label{a:A.4}

Let $A = (a_{ij})_{i, j = 1, \ldots, n}$ be a generalized Cartan matrix of finite type. The set of
roots of the Lie algebra $\gothg(A)$ contains a unique {\em maximal root\/} $\theta$ characterized
by the property that there is no root $\gamma$ of $\gothg(A)$ such that $\gamma - \theta$ is a
positive root of $\gothg(A)$. The {\em extended Cartan matrix\/} $A^{(1)} = (a_{ij})_{i, j = 0, 1,
\ldots, n}$ is obtained from $A$ by the rules $a_{00} = 2$ and
\[
a_{0i} = -2 \frac{(\theta, \alpha_i)}{(\theta, \theta)}, \qquad a_{i0} = -2 \frac{(\alpha_i,
\theta)}{(\alpha_i, \alpha_i)}
\]
for $i = 1, \ldots, n$. The matrix $A^{(1)}$ is a generalized Cartan matrix of affine type. The
Cartan subalgebra $\gothh(A)$ of $\gothg(A)$ can be naturally identified with a subalgebra of the
Cartan subalgebra $\gothh(A^{(1)})$ of $\gothg(A^{(1)})$. We identify $\gothh^*(A)$ with the
subspace of $ \gothh^*(A^{(1)})$ consisting of the elements $\gamma$ which satisfy the equations
$\gamma(h_0) = 0$ and $\gamma(d) = 0$. Hence, the set of roots $\Delta(A)$ can be considered as a
subset of $ \gothh^*(A^{(1)})$. It can be shown that
\[
\Delta(A^{(1)}) = \{ \gamma + m \delta \mid \gamma \in \Delta(A), \, m \in \bbZ \} \cup \{m \delta
\mid m \in \bbZ_+ \cup \bbZ_- \},
\]
where $\delta = \alpha_0 + \theta$. The system of positive roots is
\begin{multline*}
\Delta_+(A^{(1)}) = \{ \gamma + m \delta \mid \gamma \in \Delta_+(A), \, m \in \bbN \} \\*[.5em]
\cup \{m \delta \mid m \in \bbZ_+ \} \cup \{ (\delta - \gamma) + m \delta \mid \gamma \in
\Delta_+(A), \, m \in \bbN \}.
\end{multline*}

The affine Lie algebras $\gothg(A^{(1)})$ form the family of {\em untwisted affine Lie algebras\/}.

\subsection{Universal enveloping algebra}

Let $A = (a_{ij})$ be a generalized Cartan matrix. The universal enveloping algebra $U(\gothg(A))$
of the Kac-Moody algebra $\gothg(A)$ is a complex associative algebra with unity defined by the
same generators and relations as $\gothg(A)$ where Lie bracket is understood as commutator. Now the
Serre relations can be written as
\[
\sum_{k = 0}^{1 - a_{ij}} (-1)^k \binom{1 - a_{ij}}{k} (e_i)^{1 - a_{ij} - k} e_j (e_i)^k = 0,
\quad \sum_{k = 0}^{1 - a_{ij}} (-1)^k \binom{1 - a_{ij}}{k} (f_i)^{1 - a_{ij} - k} f_j^{} (f_i)^k = 0.
\]
It is possible to exclude the generator $d$ and consider the algebra $U(\gothg'(A))$.

\section{\texorpdfstring{Quantum groups and the universal $R$-matrix}{Quantum groups and the universal R-matrix}}

\subsection{\texorpdfstring{$q$-numbers}{q-numbers}}

We use the usual definition for $q$-number, $q$-factorial and $q$-binomial coefficient:
\begin{gather*}
[n]_q = \frac{q^n - q^{-n}}{q - q^{-1}}, \\
[n]_q! = [n]_q [n - 1]_q \ldots [1]_q, \\
\qbinom{n}{m}_q = \frac{[n]_q!}{[m]_q! [n - m]_q!}.
\end{gather*}
We assume also that the $q$-exponential is given by
\[
\exp_q(x) = \sum_{n = 0}^\infty \frac{x^n}{(n)_q!},
\]
where another $q$-deformation is used:
\[
(n)_q = \frac{q^n - 1}{q - 1}
\]
and the $q$-factorial is defined as
\[
(n)_q! = (n)_q (n - 1)_q \ldots (1)_q.
\]

\subsection{Quantum groups} \label{a:B.2}

Let $A = (a_{ij})$ be a generalized Cartan matrix of finite or affine type, $\hbar$ be an
indeterminate and $q = \exp \hbar$. The {\em quantum group\/} associated with the Lie algebra
$\gothg(A)$,\footnote{In general, one can associate the quantum group with any Kac-Moody algebra
$\gothg(A)$ with the symmetrizable Cartan matrix $A$.} is a complex associative algebra
$U_\hbar(\gothg(A))$ with unity defined by $3n$ generators $h_i$, $e_i$, $f_i$ in the finite-type
case, or by $3n + 1$ generators $h_i$, $e_i$, $f_i$, $d$ in the affine case, and by the relations
\begin{gather*}
[h_i, h_j] = 0, \\[.5em]
[h_i, e_j] = a_{ij} \, e_j, \qquad [h_i, f_j] = - a_{ij} \, f_j, \\[.5em]
[e_i, f_j] = \delta_{ij} \, \frac{q^{d_i h_i} - q^{-d_i h_i}}{q^{d_i} - q^{-d_i}}, \\[.5em]
\sum_{k = 0}^{1 - a_{ij}} (-1)^k \qbinom{1 - a_{ij}}{k}_{q^{d_i}} (e_i)^{1 - a_{ij} - k} e_j (e_i)^k = 0, \\[.5em]
\sum_{k = 0}^{1 - a_{ij}} (-1)^k \qbinom{1 - a_{ij}}{k}_{q^{d_i}} (f_i)^{1 - a_{ij} - k} f_j^{} (f_i)^k = 0, \\[.5em]
[d, e_i] = \delta_{0i} e_i, \qquad [d, f_i] = - \delta_{0i} f_i, \qquad [d, h_i] = 0.
\end{gather*}
It is worth to note that we consider the elements of $U_\hbar(\gothg(A))$ as formal power series in
$\hbar$. If we exclude the generator $d$ from our consideration, we will obtain an algebra denoted
by $U_\hbar(\gothg'(A))$.

We define the quantum analog of the Cartan anti-involution by the relations
\begin{equation}
\omega(e_i) = f_i, \qquad \omega(f_i) = e_i, \qquad \omega(h_i) = h_i, \qquad \omega(d) = d
\label{ci}
\end{equation}
together with the rule $\omega(\hbar) = - \hbar$ implying that $\omega(q) = q^{-1}$.

The algebras $U_\hbar(\gothg(A))$ and $U_\hbar(\gothg'(A))$ are Hopf algebras with
co-multiplication defined by the relations
\begin{gather*}
\Delta(h_i) = h_i \otimes 1 + 1 \otimes h_i, \\
\Delta(e_i) = e_i \otimes 1 + q^{-d_i h_i} \otimes e_i, \qquad \Delta(f_i)
= f_i \otimes q^{d_i h_i} + 1 \otimes f_i, \\
\Delta(d) = d \otimes 1 + 1 \otimes d.
\end{gather*}

The Cartan subalgebra $\gothh(A)$ can be naturally identified with the vector subspace of
$U_\hbar(\gothg(A))$ formed by linear combinations of the generators $h_i$ and $d$, and we say that
an element $a \in U_\hbar(\gothg(A))$ is a {\em root vector\/} corresponding to a root $\gamma \in
\gothh^*(A)$ if $\gamma \ne 0$ and
\[
[h, a] = \gamma(h) a
\]
for all $h \in \gothh(A)$. It is evident that actually $\gamma \in Q(A)$. Let $a$ and $b$ be root
vectors corresponding to roots $\alpha$ and $\beta$ respectively. We define the {\em
$q$-commutator\/} $[a, b]_q$ of $a$ and $b$ as
\[
[a, b]_q = a b - q^{(\alpha, \beta)} b a
\]
if $\alpha, \beta \in Q_+(A)$, as
\[
[a, b]_q = a b - q^{-(\alpha, \beta)} b a
\]
if $\alpha, \beta \in Q_-(A)$, and as the usual commutator if $\alpha \in Q_+(A)$ and $\beta \in
Q_-(A)$, or $\alpha \in Q_-(A)$ and $\beta \in Q_+(A)$.

\subsection{Symmetric group and tensor products}

Let $\calA$ be an associative unital algebra. Define an action of the symmetric group S$_n$ on the
tensor product $\calA^{\otimes n}$ in the following way. Let ${s} \in \mathrm S_n$, define a linear
operator $\Pi_{s}$ on $\calA^{\otimes n}$ by the equation
\[
\Pi_{s}(a_1 \otimes a_2 \otimes \ldots \otimes a_n) = a_{s^{-1}(1)} \otimes a_{s^{-1}(2)} \otimes
\ldots \otimes a_{s^{-1}(n)}.
\]
The set of operators $\Pi_{s}$ define a left action of S$_n$ on $\calA^{\otimes n}$, i. e. we have
\[
\Pi_{s} \circ \Pi_{t} = \Pi_{s t}
\]
for all ${s, t} \in \mathrm S_n$.

Let $M$ be an element of the tensor product $\calA^{\otimes k}$, where $k \le n$. Denote by $M_{12
\ldots k}$ the element of $\calA^{\otimes n}$ defined as
\[
M_{12 \ldots k} = M \otimes \underbrace{1 \otimes 1 \otimes \ldots \otimes 1}_{n-k}.
\]
Let $i_1, i_2, \ldots, i_k$ are distinct integers in the range from $1$ to $n$. We define
\[
M_{i_1 i_2 \ldots i_k} = \Pi_{t}(M_{12 \ldots k}),
\]
where $t$ is any element of S$_n$ such that
\[
{t}(1) = i_1, \qquad {t}(2) = i_2, \qquad \ldots, \qquad {t}(k) = i_k.
\]
Here for any ${s} \in \mathrm S_n$ we have
\[
\Pi_{s}(M_{i_1 i_2 \ldots i_k}) = M_{s(i_1) s(i_2) \ldots s(i_k)}.
\]

Now let $V$ be a vector space and $\calA = \End(V)$. Given ${s} \in \mathrm S_n$, we define a
linear operator $P_{s}$ on $V^{\otimes n}$ by
\[
P_{s}(v_1 \otimes v_2 \otimes \ldots \otimes v_n) = v_{s^{-1}(1)} \otimes v_{s^{-1}(2)} \otimes
\ldots \otimes v_{s^{-1}(n)}.
\]
The set of operators $P_{s}$ define a left action of S$_n$ in $V^{\otimes n}$. We have the relation
\[
\Pi_{s} (M_{i_1 i_2 \ldots i_k}) = P^{}_{s} M_{i_1 i_2 \ldots i_k} P^{-1}_{s}
\]
which implies the equation
\begin{equation}
P_{s} M_{i_1 i_2 \ldots i_k} = M_{s(i_1) s(i_2) \ldots s(i_k)} P_{s}. \label{PMP}
\end{equation}

If $s$ is a transposition $(i \, j)$ we write $\Pi_{ij}$ and $P_{ij}$ instead of $\Pi_{s}$ and
$P_{s}$, respectively. If $n = 2$ we denote $\Pi = \Pi_{12}$ and $P = P_{12}$.

\subsection{\texorpdfstring{Universal $R$-matrix}{Universal R-matrix}}

Let $\calA$ be a Hopf algebra with co-multiplication $\Delta$. One can show that $\calA$ is a Hopf
algebra with respect to the opposite co-multiplication $\Delta^{\mathrm{op}} = \Pi \circ \Delta$.
The Hopf algebra $\calA$ is said to be {\em almost co-commutative\/} if there exists an invertible
element $\calR \in \calA \otimes \calA$ such that
\[
\Delta^{\mathrm{op}}(a) = \calR \Delta(a) \calR^{-1}
\]
for all $a \in \calA$. An almost co-commutative Hopf algebra $\calA$ is called {\em
quasi-triangular\/} if
\[
\Delta \otimes \id (\calR) = \calR_{13} \calR_{23}, \qquad \id \otimes \Delta(\calR) = \calR_{13} \calR_{12}.
\]
In this case the element $\calR$ is called the {\em universal $R$-matrix\/}. One can show that the
universal $R$-matrix satisfies the Yang-Baxter equation
\begin{equation}
\calR_{12} \, \calR_{13} \, \calR_{23} = \calR_{23} \, \calR_{13} \, \calR_{12}.
\label{cRRR}
\end{equation}

\subsection{\texorpdfstring{$R$-matrices}{R-matrices}}

Assume that for any $\zeta \in \bbC^\times$ there is given a homomorphism $\varphi_\zeta$ from
$\calA$ to a unital associative algebra $\calB$. We denote
\[
R(\zeta_1, \zeta_2) = \varphi_{\zeta_1} \otimes \varphi_{\zeta_2} (\calR).
\]
By definition, $R(\zeta_1, \zeta_2)$ for any $\zeta_1$ and $\zeta_2$ is an element of the algebra
$\calB \otimes \calB$. It follows from equation (\ref{cRRR}) that in $\calB \otimes \calB \otimes
\calB$ we have
\begin{equation}
R_{12}(\zeta_1, \zeta_2) \, R_{13}(\zeta_1, \zeta_3) \, R_{23} (\zeta_2, \zeta_3)
= R_{23}(\zeta_2, \zeta_3) \, R_{13}(\zeta_1, \zeta_3) \, R_{12}(\zeta_1, \zeta_2).
\label{RRRzz}
\end{equation}
Assume that $R(\zeta_1, \zeta_2)$ actually depends on $\zeta_1 / \zeta_2$ only. In this case we define
\begin{equation}
R(\zeta_{12})  = \varphi_{\zeta_1} \otimes \varphi_{\zeta_2} (\calR) \label{rz}
\end{equation}
and call $R(\zeta)$ an $R$-matrix. Here and below we denote $\zeta_{ij} = \zeta_i / \zeta_j$. Now
(\ref{RRRzz}) takes the form
\[
R_{12}(\zeta_{12}) \, R_{13}(\zeta_{13}) \, R_{23} (\zeta_{23}) = R_{23}(\zeta_{23}) \,
R_{13}(\zeta_{13}) \, R_{12}(\zeta_{12}).
\]

Consider the case where $\varphi$ is a representation of the algebra $\calA$ in some vector space
$V$. Here we can assume that $\calB = \End(V)$, and that $R(\zeta) \in \End(V) \otimes \End(V)
\simeq \End(V \otimes V)$. Using the definition
\[
\check R(\zeta) = P R(\zeta)
\]
and the relation (\ref{PMP}), we rewrite the left-hand side of the Yang--Baxter equation in the
following way:
\begin{multline*}
R_{12}(\zeta_{12}) R_{13}(\zeta_{13}) R_{23}(\zeta_{23}) \\
= P_{12} \check R_{12}(\zeta_{12}) P_{13} \check R_{13}(\zeta_{13}) P_{23} \check R_{23}(\zeta_{23})
= P_{12} P_{13} P_{23} \check R_{23}(\zeta_{12}) \check R_{12}(\zeta_{13}) \check R_{23}(\zeta_{23}).
\end{multline*}
Similarly, we rewrite the right-hand side as
\[
R_{23}(\zeta_{23}) R_{13}(\zeta_{13}) R_{12}(\zeta_{12}) = P_{23} P_{13} P_{12} \check
R_{12}(\zeta_{23}) \check R_{23}(\zeta_{13}) \check R_{12}(\zeta_{12}).
\]
It is not difficult to verify that
\[
P_{12} P_{13} P_{23} = P_{23} P_{13} P_{12},
\]
therefore, the Yang--Baxter equation is equivalent to the equation
\[
\check R_{23}(\zeta_{12}) \check R_{12}(\zeta_{13}) \check R_{23}(\zeta_{23}) = \check
R_{12}(\zeta_{23}) \check R_{23}(\zeta_{13}) \check R_{12}(\zeta_{12}).
\]
This equation can also be written as
\[
(1 \otimes \check R(\zeta_{12})) (\check R(\zeta_{13}) \otimes 1) (1 \otimes \check R(\zeta_{23}))
= (\check R(\zeta_{23}) \otimes 1) (1 \otimes \check R(\zeta_{13})) (\check R(\zeta_{12}) \otimes
1).
\]

In a similar way one can show that in terms of
\[
\hat R(\zeta) = R(\zeta) P
\]
the Yang--Baxter equation has the form
\[
(\hat R(\zeta_{12}) \otimes 1) (1 \otimes \hat R(\zeta_{13})) (\hat R(\zeta_{23}) \otimes 1) = (1
\otimes \hat R(\zeta_{23})) (\hat R(\zeta_{13}) \otimes 1) (1 \otimes \hat R(\zeta_{12})).
\]

Let $e_a$ be a basis of $V$. Define endomorphisms $E_{ab} \in \End(V)$ with the help of the
equations
\[
E_{ab}(e_c) = e_a \delta_{bc}.
\]
It follows from this definition that
\begin{equation}
E_{ab} \circ E_{cd} = \delta_{bc} E_{ad}. \label{EEE}
\end{equation}
For any $M \in \End(V)$ we have
\[
M(e_b) = \sum_a e_a M_{ab}
\]
for appropriate numbers $M_{ab}$, and one can verify that
\[
M = \sum_{a, b} E_{ab} M_{ab}.
\]
It is easy to show that the endomorphisms $E_{ab}$ are linearly independent. Thus, they form a
basis of the vector space $\End(V)$.

One can be convinced that the endomorphisms $E_{ac} \otimes E_{bd}$ form a basis of the vector
space $\End(V \otimes V)$ and we have
\[
R(\zeta) = \sum_{a,b,c,d} E_{ac} \otimes E_{bd} \, R_{ab, cd}(\zeta),
\]
where $R_{ab, cd}(\zeta)$ are some functions of $\zeta$. Usually, one identifies $R(\zeta)$ with a
matrix-valued function formed by the functions $R_{ab, cd}(\zeta)$, hence the name $R$-matrix.

Defining functions $\hat R_{ab, cd}(\zeta)$ by
\[
\hat R(\zeta) = \sum_{a,b,c,d} E_{ac} \otimes E_{bd} \, \hat R_{ab, cd}(\zeta),
\]
we see that
\begin{equation}
\hat R_{ab, cd}(\zeta) = R_{ab, dc}(\zeta).
\label{hrabcd}
\end{equation}
Similarly, defining functions $\check R_{ab, cd}(\zeta)$ by
\[
\check R(\zeta) = \sum_{a,b,c,d} E_{ac} \otimes E_{bd} \, \check R_{ab, cd}(\zeta),
\]
we see that
\[
\check R_{ab, cd}(\zeta) = R_{ba, cd}(\zeta).
\]

\subsection{\texorpdfstring{$L$-operators}{L-operators}} \label{a:B.6}

For any $\zeta \in \bbC^\times$ let $\varphi_\zeta$ be a homomorphism  from $\calA$ to a unital
associative algebra $\calB$ and $\chi_\zeta$ a homomorphism from $\calA$ to a unital associative
algebra $\calC$. By $\hat L(\zeta_1, \zeta_2)$ we denote  an element of $\calC \otimes \calB$
defined as
\[
\hat L(\zeta_1, \zeta_2) = \chi_{\zeta_1} \otimes \varphi_{\zeta_2} (\calR).
\]
If $\hat L(\zeta_1, \zeta_2)$ only depends on $\zeta_1/\zeta_2$, we define
\begin{equation}
\hat L(\zeta_{12}) = \chi_{\zeta_1} \otimes \varphi_{\zeta_2} (\calR)
\label{hlz}
\end{equation}
and call the element $\hat L(\zeta)$ an {\em $L$-operator\/}. It follows from (\ref{cRRR}) that the
$R$-matrix and $L$-operator satisfy the equation
\begin{equation}
R_{23}(\zeta_{12}) \hat L_{13}(\zeta_1) \hat L_{12}(\zeta_2) = \hat L_{12}(\zeta_2) \hat
L_{13}(\zeta_1) R_{23} (\zeta_{12}), \label{RLL}
\end{equation}
where $R_{23}(\zeta)$, $\hat L_{13}(\zeta)$ and $\hat L_{12}(\zeta)$ are elements of $\calC \otimes
\calB \otimes \calB$ defined in an evident way.

Assume again that for any $\zeta \in \bbC^\times$ the homomorphism $\varphi_\zeta$ is a
representation of $\calA$ in a vector space $V$. It is clear that in this case one has
\[
\hat L(\zeta) = \sum_{a, b} \hat L_{ab} \otimes E_{ab},
\]
where $\hat L_{ab}(\zeta)$ are some $\calC$-valued functions of $\zeta$. Using equation (\ref{EEE})
one can show that equation (\ref{RLL}) is equivalent to the system of equations
\begin{equation}
\sum_{e,f} R_{ab, ef}(\zeta_{12}) \, \hat L_{fc}(\zeta_1) \, \hat L_{ed}(\zeta_2)
= \sum_{e,f} \hat L_{ae}(\zeta_2) \, \hat L_{bf}(\zeta_1) \, R_{ef, dc}(\zeta_{12}).
\label{RLLM}
\end{equation}
Taking into account (\ref{hrabcd}), we rewrite equations (\ref{RLLM}) as
\[
\sum_{e,f} \hat R_{ab, ef}(\zeta_{12}) \, \hat L_{ec}(\zeta_1) \, \hat L_{fd}(\zeta_2)
= \sum_{e,f} \hat L_{ae}(\zeta_2) \, \hat L_{bf}(\zeta_1) \, \hat R_{ef, cd}(\zeta_{12}).
\]
Identifying $\hat R(\zeta)$ with a matrix-valued function formed by the functions $\hat R_{ab,
cd}(\zeta)$ and $\hat L(\zeta)$ with a matrix-valued function formed by the algebra-valued
functions $\hat L_{ab}(\zeta)$, we have
\begin{equation}
\hat R(\zeta_{12}) (\hat L(\zeta_1) \times \hat L(\zeta_2)) = (\hat L(\zeta_2) \times \hat
L(\zeta_1)) \hat R(\zeta_{12}), \label{hrll}
\end{equation}
where $\times$ means the generalization of the Kronecker product to the matrices with algebra-valued
entries. Explicitly, if $M = (M_{ab})$, where $M_{ab}$ are elements of some algebra, and $N =
(N_{ij})$, where $N_{ij}$ are elements of the same algebra, then $M \times N = ((M \times N)_{ai,
bj})$, where
\begin{equation}
(M \times N)_{ai, bj} = M_{ab} N_{ij}.
\label{akp}
\end{equation}
We follow here the notation used in \cite{ChaPre95}. We call a matrix-valued function $\hat
L(\zeta)$ formed by algebra-valued functions $\hat L_{ab}(\zeta)$ an {\em $L$-operator of type
$\hat L$\/} if it satisfies the relation (\ref{hrll}).

One can define another type of $L$-operator,
\begin{equation}
\check L(\zeta_{12}) = \varphi_{\zeta_1} \otimes \psi_{\zeta_2}(\calR),
\label{clz}
\end{equation}
where $\psi_\zeta$ for any $\zeta \in \bbC^\times$ is a homomorphism from $\calA$ to a unital
associative algebra $\calD$. In the case where for any $\zeta \in \bbC^\times$ the homomorphism
$\varphi_\zeta$ is a representation of $\calA$ in the vector space $V$, one can write
\[
\check L(\zeta) = \sum_{a,b} E_{ab} \otimes \check L_{ab}(\zeta).
\]
Identifying $\check R(\zeta)$ with a matrix-valued function formed by the functions $\check R_{ab,
cd}(\zeta) = R_{ba, cd}(\zeta)$ and $\check L(\zeta)$ with a matrix-valued function formed by the
functions $\check L_{ab}(\zeta)$, we obtain
\begin{equation}
\check R(\zeta_{12}) (\check L(\zeta_1) \times \check L(\zeta_2)) = (\check L(\zeta_2) \times \check
L(\zeta_1)) \check R(\zeta_{12}). \label{crll}
\end{equation}
We call a matrix-valued function $\check L(\zeta)$ formed by algebra-valued functions $\check
L_{ab}(\zeta)$ an {\em $L$-operator of type $\check L$\/} if it satisfies the relation
(\ref{crll}).

\subsection{\texorpdfstring{Khoroshkin--Tolstoy construction}{Khoroshkin-Tolstoy construction}}
\label{ktc}

Khoroshkin and Tolstoy proposed a procedure to construct the universal $R$-matrix for the quantum
groups $U_\hbar(\gothg'(A^{(1)}))$ associated to untwisted affine Lie algebras \cite{TolKho92}. An
example of an extension of the procedure to the case of twisted affine Lie algebras is considered
in \cite{KhoTol92}. For simplicity we consider the case when the initial Cartan matrix $A$ is
symmetric. In this case the extended Cartan matrix $A^{(1)}$ is also symmetric.

The first step of the procedure is to choose a special ordering of the positive roots of $\gothg(A^{(1)})$.

We say that the system of positive roots $\Delta_+(\gothg(A^{(1)}))$ is supplied with a {\em normal
order\/} if its roots are totally ordered in such a way that

\begin{itemize}
\item[(i)] all multiple roots follow each other in an arbitrary order;
\item[(ii)] each non-simple root $\alpha + \beta$, where $\alpha$ is not proportional
to $\beta$ is placed between $\alpha$ and $\beta$.
\end{itemize}

We fix some normal order of $\Delta_+(A^{(1)})$ which satisfies the additional condition that for
any root $\gamma \in \Delta_+(A)$ one has
\begin{equation}
\gamma + m \delta \prec k \delta \prec (\delta - \gamma) + \ell \delta.
\label{noc}
\end{equation}

The second step of the procedure is to construct the root vectors corresponding to the positive
roots of $\gothg(A^{(1)})$ from the root vectors corresponding to the simple positive roots
$e_{\alpha_0} = e_{\delta - \theta}$ and $e_{\alpha_i}$.

First, we construct the root vectors corresponding to the roots $\gamma$ and $\delta - \gamma$,
$\gamma \in \Delta_+(A)$. Here if $\gamma = \alpha + \beta$, $\alpha \prec \gamma \prec \beta$, and
there are no roots $\alpha'$ and $\beta'$ between $\alpha$ and $\beta$ such that $\gamma = \alpha'
+ \beta'$, we define
\begin{equation}
e_\gamma = [e_\alpha, e_\beta]_q.
\label{e1}
\end{equation}
Then we take
\begin{equation}
e_{\delta - \gamma} = [e_{\theta - \gamma}, e_{\delta - \theta}]_q.
\label{e2}
\end{equation}
The root vectors corresponding to the root $\delta$ are indexed by the elements of $\Delta_+(A)$
and defined by the relation
\begin{equation}
e'_{\delta, \, \gamma} = [e_\gamma, \, e_{\delta - \gamma}]_q.
\label{e3}
\end{equation}
The remaining definitions are
\begin{gather}
e_{\gamma + m \delta} = [(\gamma, \gamma)]_q^{-1} [e_{\gamma + (m - 1)\delta}, \, e'_{\delta, \,
\gamma}]_q,
\label{e4} \\
e_{(\delta - \gamma) + m \delta} = [(\gamma, \gamma)]_q^{-1} [e'_{\delta, \, \gamma}, \, e_{(\delta
- \gamma) + (m - 1)\delta}]_q,
\label{e5} \\
e'_{m \delta, \, \gamma} = [e_{\gamma + (m - 1)\delta}, \, e_{\delta - \gamma}]_q.
\label{e6}
\end{gather}
Actually from $e'_{m \delta, \, \gamma}$ we have to proceed to $e_{m \delta, \, \gamma}$ defined by
the equation
\begin{equation}
(q - q^{-1}) e_{\delta, \gamma}(x) = \log(1 + (q - q^{-1}) e'_{\delta, \, \gamma}(x)),
\label{e7}
\end{equation}
where
\[
e'_{\delta, \, \gamma}(x) = \sum_{m=1}^\infty e'_{m \delta, \, \gamma} x^{-m}, \qquad e_{\delta, \,
\gamma}(x) = \sum_{m=1}^\infty e_{m \delta, \, \gamma} x^{-m}.
\]

The root vectors, corresponding to the negative roots, are constructed with the help of the Cartan
anti-involution:
\begin{gather}
f'_{m \delta, \, \gamma} = \omega(e'_{m \delta, \, \gamma}), \qquad f_{m \delta, \, \gamma}
= \omega(e_{m \delta, \, \gamma}), \label{f1} \\
f_{\gamma + m \delta} = \omega(e_{\gamma + m \delta}),  \qquad f_{(\delta - \gamma) + m \delta}
= \omega(e_{(\delta - \gamma) + m \delta}).
\label{f2}
\end{gather}
In particular, for a composite root $\gamma \in \Delta_+(A)$, such that $\gamma = \alpha + \beta$
with $\alpha \prec \gamma \prec \beta$, we have
\begin{equation}
f_\gamma = [f_\beta, f_\alpha]_q.
\label{fab}
\end{equation}
Further, for any $\gamma \in \Delta_+(A)$ the analog of the definition (\ref{e2}) is
\begin{equation}
f_{\delta - \gamma} = [f_{\delta - \theta}, f_{\theta - \gamma}]_q.
\label{fdg}
\end{equation}
Finally, one can be convinced that
\begin{gather}
f_{\gamma + m \delta} = [(\gamma, \gamma)]_q^{-1} [f'_{\delta, \, \gamma}, \, f_{\gamma + (m -
1)\delta}]_q,
\label{f4} \\
f_{(\delta - \gamma) + m \delta} = [(\gamma, \gamma)]_q^{-1} [f_{(\delta - \gamma)
+ (m - 1)\delta}, \, f'_{\delta, \, \gamma}]_q,
\label{f5} \\
f'_{m \delta, \, \gamma} = [f_{\delta - \gamma}, \, f_{\gamma + (m - 1)\delta}]_q \label{f6}
\end{gather}
for all $m > 0$, and that
\begin{equation}
-(q - q^{-1}) f_{\delta, \gamma}(x) = \log(1 - (q - q^{-1}) f'_{\delta, \, \gamma}(x)),
\label{f7}
\end{equation}
where
\[
f'_{\delta, \, \gamma}(x) = \sum_{m=1}^\infty f'_{m \delta, \, \gamma} x^{-m}, \qquad f_{\delta, \,
\gamma}(x) = \sum_{m=1}^\infty f_{m \delta, \, \gamma} x^{-m}.
\]

The expression for the universal $R$-matrix obtained by Khoroshkin and Tolstoy has the form
\[
\calR = \calR_{\prec \delta} \, \calR_{\sim \delta} \, \calR_{\succ \delta} \, \calK.
\]
The first factor is the product over $\gamma \in \Delta_+(A)$ and $m \in \bbN$ of the
$q$-exponentials
\begin{equation}
\calR_{\gamma, \, m} = \exp_{q^{-(\gamma, \gamma)}} \left( (q - q^{-1})
\, s^{-1}_{m, \, \gamma} \, e_{\gamma + m \delta} \otimes f_{\gamma + m \delta} \right).
\label{rgm}
\end{equation}
Here the quantities $s_{m, \, \gamma}$ are determined by the relation
\[
[e_{\gamma + m \delta}, \, f_{\gamma + m \delta}] = s_{m, \, \gamma} \, \frac{q^{h_{\gamma + m
\delta}} - q^{-h_{\gamma + m \delta}}}{q - q^{-1}},
\]
where $h_{\gamma + m \delta} = \sum_i k_i h_i$ if $\gamma + m \delta = \sum_i k_i \alpha_i$. The
order of the factors in $\calR_{\prec \delta}$ coincides with the chosen normal order of the roots
$\gamma + m \delta$. The second factor is
\begin{equation}
\calR_{\sim \delta} = \exp \left( (q - q^{-1}) \sum_{m \in \bbZ_+} \sum_{i, j
= 1}^r u_{m, \, ij} \, e_{m \delta, \, \alpha_i} \otimes f_{m \delta, \, \alpha_j} \right),
\label{rpd}
\end{equation}
where for each $m \in \bbZ_+$ the quantities $u_{m, \, ij}$ are the matrix elements of the matrix
$u_m$ inverse to the matrix $t_m$ with the matrix elements
\[
t_{m, \, ij} = (-1)^{m(1 - \delta_{ij})} m^{-1} [m a_{ij}]_q,
\]
entering the commutation relations
\[
[e_{\alpha_i + m \delta}, \, e_{n \delta, \alpha_j}]_q = t_{n, i j} \, e_{\alpha_i + (m + n)\delta},
\]
and $r$ is the rank of the Lie algebra $\gothg(A)$. The definition of the factor $\calR_{\succ
\delta}$ is similar to the definition of the factor $\calR_{\prec \delta}$. It is the product over
$\gamma \in \Delta_+(A)$ and $m \in \bbN$ of the $q$-exponentials
\begin{equation}
\calR_{\delta - \gamma, \, m} = \exp_{q^{-(\gamma, \gamma)}} \left( (q - q^{-1}) \, s^{-1}_{m, \,
\delta - \gamma} \, e_{(\delta - \gamma) + m \delta} \otimes f_{(\delta - \gamma) + m \delta}
\right). \label{rdmgm}
\end{equation}
The quantities $s_{m, \, \delta - \gamma}$ are determined by the relation
\[
[e_{(\delta - \gamma) + m \delta}, \, f_{(\delta - \gamma) + m \delta}] = s_{m, \, \delta - \gamma}
\,  \frac{q^{h_{(\delta - \gamma) + m \delta}} - q^{-h_{(\delta - \gamma) + m \delta}}}{q -
q^{-1}},
\]
where $h_{(\delta - \gamma) + m \delta} = \sum_i k_i h_i$ if $(\delta - \gamma) + m \delta = \sum_i
k_i \alpha_i$. The order coincides with the chosen normal order of the roots $(\delta - \gamma) + m
\delta$. For the factor $\calK$ we have the expression
\begin{equation}
\calK = \exp \left( \hbar \sum_{i, j = 1}^r (b_{ij} \, h_{\alpha_i} \otimes h_{\alpha_j}) \right),
\label{k}
\end{equation}
where $b_{ij}$ are the matrix elements of the matrix inverse to the Cartan matrix $A = (a_{ij})$.

We determine that the universal $R$-matrix in the case under consideration is an element of
$U_\hbar(\gothb'_+(A^{(1)})) \otimes U_\hbar(\gothb'_-(A^{(1)}))$, where
$U_\hbar(\gothb'_+(A^{(1)}))$ and $U_\hbar(\gothb'_-(A^{(1)}))$ are the associative unital algebras
defined by the generators $e_i$, $h_i$ and $f_i$, $h_i$, respectively. It means, in particular,
that in order to define an $L$-operator of type $\hat L$ it is enough to assume that $\chi_\zeta$
for any $\zeta \in \bbC^\times$ is a homomorphism from $U_\hbar(\gothb'_+(A^{(1)}))$ to some unital
associative algebra. Similarly, to define an $L$-operator of type $\check L$ it is enough to assume
that $\psi_\zeta$ for any $\zeta \in \bbC^\times$ is a homomorphism from
$U_\hbar(\gothb'_-(A^{(1)}))$ to some unital associative algebra.

To construct an $R$-matrix we should define the corresponding homomorphisms $\varphi_\zeta$. In the
case under consideration, the simplest way to do this is to start with a $\zeta$-indepen\-dent
homomorphism $\varphi$ and then define $\varphi_\zeta$ by the relations
\begin{gather}
\varphi_\zeta(h_{\alpha_i}) = \varphi(h_{\alpha_i}),
\label{phiz1} \\
\varphi_\zeta(e_{\alpha_i}) = \zeta^{s_i} \varphi(e_{\alpha_i}), \qquad \varphi_\zeta(f_{\alpha_i})
= \zeta^{-s_i} \varphi(f_{\alpha_i}), \label{phiz2}
\end{gather}
where $s_i$ are some integers. It is not difficult to understand that in this case the $R$-matrix
$R(\zeta_1, \zeta_2)$, defined by equation (\ref{rz}), depends only on $\zeta_1/\zeta_2$. In the
same way one can define the homomorphisms $\chi_\zeta$ and $\psi_\zeta$. The corresponding
$L$-operators $\hat L(\zeta_1, \zeta_2)$ and $\check L(\zeta_1, \zeta_2)$, defined by equations
(\ref{hlz}) and (\ref{clz}), respectively, also depend only on $\zeta_{12}$.

\vskip3mm {\em Acknowledgements.\/} We are grateful to Professor V. V. Bazhanov for discussions.
This work was supported in part by the Volkswagen Foundation and the joint DFG-RFBR grant No.
08-01-91953.

\bibliographystyle{amsrusplain}

\bibliography{UniversalRMatrix}

\end{document}